\documentclass{aa}
\usepackage{psfrag,graphicx}
\usepackage{txfonts}
\usepackage{natbib}
\usepackage{ulem}
\usepackage{color}

\begin{document}
\bibliographystyle{aa}

   \title{The Thermodynamics of Molecular Cloud Fragmentation}
   \subtitle{Star Formation Under Non-Milky Way Conditions}

   \author{S. Hocuk \inst{1}
          \and
          M. Spaans \inst{1}
          }
   \offprints{S. Hocuk}
   \institute{Kapteyn Astronomical Institute, University of Groningen,
              P.~O.~Box 800, 9700 AV Groningen \\
              \email{seyit@astro.rug.nl, spaans@astro.rug.nl}
             }

   \titlerunning{Fragmenting Clouds}
   \authorrunning{S.~Hocuk \& M.~Spaans}
   \date{Received \today}

\abstract
{Properties of candidate stars, forming out of molecular clouds, depend on the ambient conditions of the parent cloud. We present a series of 2D and 3D simulations of fragmentation of molecular clouds in starburst regions as well as clouds under conditions in dwarf galaxies, leading to the formation of protostellar cores.}
{We explore in particular the metallicity dependence of molecular cloud fragmentation and the possible variations in the dense core mass function, as the expression of a multi-phase ISM, due to dynamic and thermodynamic effects in starburst and metal-poor environments.}
{To study the level of fragmentation during the collapse, the adaptive mesh refinement code FLASH is used. With this code, including self-gravity, thermal balance, turbulence and shocks, collapse is simulated with four different metallicity dependent cooling functions. Turbulent and rotational energies are considered as well. During the simulations, number densities of 10$^8$-10$^9 \rm ~cm^{-3}$ are reached. The influences of dust and cosmic ray heating are investigated and a comparison to isothermal cases is made.}
{The presented results indicate that fragmentation increases with metallicity, while cosmic ray and gas-grain collisional heating counteract this. We also find that modest rotation and turbulence can impact the cloud evolution as far as fragmentation is concerned. In this light, we conclude that radiative feedback, in starburst regions, will inhibit fragmentation, while low-metallicity dwarfs also should enjoy a star formation mode in which fragmentation is suppressed.}
{}

\maketitle

\section{Introduction}
The initial mass function (IMF) in our local neighborhood is observed to be a power-law function, nicely following a Salpeter slope. Recent studies \citep{2005ASSL..327...89F, 2005Natur.434..192F, 2007AAS...210.1601S, 2008ApJ...681..365E} confirm this universality at low and high masses. When we turn to observations of more radical environments like the Galaxy center, a 'universal' shape is again confirmed and only hints for different IMFs appear. Measurements of abundance patterns \citep{2007A&A...467..117B, 2008IAUS..245..339C} indicate a flatter IMF. Observations of the Arches cluster initially showed a shallow IMF \citep{1999ASPC..186..329F, 2002A&A...394..459S} and a turn-over mass, and thus a typical mass, at a few solar masses \citep[6-7M$_{\odot}$,][]{2005ApJ...628L.113S}. However, \cite{2006ApJ...653L.113K} do detect low mass stars with their deep photometry and merely find the presence of a local bump at $\sim$6.3M$_{\odot}$ and a somewhat shallower IMF ($\Gamma = -1.0$ to $-1.1$). \cite{2007MNRAS.381L..40D} confirm these results and claim a top-heavy IMF. Again, there appear to be perfectly reasonable explanations for the higher mass stars detected in the Arches cluster as \cite{2007MNRAS.378L..29P} argue with their idea of a still collapsing cluster core. Of course, the idea that the conditions and the environment play an important role in shaping the IMF is worth studying. Looking at the mass-to-light ratios of ultra-compact dwarf galaxies, \cite{2009arXiv0901.0915D} conclude from their models that it is most likely due to a top-heavy stellar IMF. If so, then they attribute this to feedback effects induced by massive stars. Still, these studies do not benefit from resolved stellar population data. One can associate the conditions and the environment of dwarf galaxies with those in the early Universe \citep{2008IAUS..255..226H, 2009MNRAS.tmpL.207S}. It is also believed that the primordial, high redshift and zero metallicity, IMF must have been top-heavy, $\sim$100M$_{\odot}$ \citep{2000ApJ...540...39A, 2001ApJ...561L..55O, 2002ApJ...564...23B, 2002Sci...295...93A, 2003ApJ...589..677O, 2008arXiv0810.1867J}. Hence, it appears that the IMF is universal, but that it is worthwhile to explore strong environmental (feedback) influences, e.g., starburst and dwarf galaxies, to understand the IMF better \citep{2007MNRAS.374L..29K}. Since star formation, fragmentation and the IMF all depend on initial conditions \citep{2003ApJ...596..253S, 2008IAUS..255...33G}, the question is how strong the impact of these initial conditions is. Also, how different should the environment be, or how strong the feedback, to cause a significant deviation from a Salpeter IMF?

Molecular clouds are the birth grounds of stars. These clouds are formed (rapidly) when interstellar gas and dust gather and cool down under the influence of hydrodynamics, thermodynamics and gravity \citep{1999ApJ...527..285B, 2001ApJ...562..852H, 2005ApJ...620..786K, 2008ApJ...689..290H, 2008MNRAS.385.1893D, 2008MNRAS.391..844D}. Molecules form in the gas phase and on dust grains, and allow low temperatures to be reached \citep{2005ApJ...626..627O, 2005JPhCS...6..155C, 2009A&A...496..365C, 2009arXiv0903.3120D}. Dense regions, like the center of the cloud, are able to form many stars, with a modest efficiency \citep[$\sim$5\%;][]{2000ApJ...539..342E, 2005MNRAS.359..809C}. The distribution of the dense regions is thought to be the precursor of a stellar initial mass function \citep{1998A&A...336..150M, 2001ASPC..243..301M, 2008A&A...477..823G, 2008MNRAS.391..205S}, and perhaps the stellar IMF is directly linked to this core mass function as it seems for the Pipe nebula \citep{2008ApJ...672..410L}. Following a typical molecular cloud from its infancy to a mature state, for different starting conditions, gives insight into the processes that determine stellar masses and is the focus of this work. Specifically, we consider low metallicity environments (dwarfs) and warm dusty systems (starbursts).

In the next section we present the initial conditions used for our 38 simulations and present the results in the following section. Twenty of these simulations form the basis of this research. They comprise 4 different metallicities with 5 combinations of turbulent and rotational energies each. We add to these basis simulations comparison studies that include extra heating sources and cases where the primary heating and cooling terms are removed.

\section{Simulations}
\subsection{Numerical Model and Simulation Setup}
In order to model a collapsing molecular cloud, we solve the hydrodynamical equations using the adaptive mesh refinement (AMR) code FLASH \citep{2000ApJS..131..273F}. FLASH uses the PARAMESH library \citep{1999AAS...195.4203O, 2000CoPhC.126..330M} for grid refinement and parallelization. Hydrodynamic equations are solved using the piecewise parabolic method (PPM, \cite{1984JCoPh..54..174C}).

For our 2D simulations we use a grid with a maximum resolution that amounts to 4096$^2$ cells when completely refined, which is a refinement level of 10 in FLASH terms. To minimize computational demands, we utilize the power of adaptive meshes by initializing the simulation at refinement level 5, i.e., 128$^2$ cells, and refine with a density threshold of $n = 600 \rm ~cm^{-3}$ in order not to violate the Truelove criteria \citep{1997ApJ...489L.179T}. This density was found to be the optimal balance between following the physics of cloud collapse and computational demand. The most dynamic `central' area, however, always starts with the highest refinement to maximize precision and minimize resolution dependent effects in this region. The border resolution between the refinement levels is set-up in such a way that it is not a sharp transition between the minimum and the maximum refinement.

We are aware of the problems that might arise if resolution is not sufficient. \cite{1997ApJ...489L.179T} state that the Jeans length should at least be resolved by 4 cells, i.e., N$_j>4$ in order to ensure that the collapse is of a physical rather than numerical nature. Another pitfall is that the transfer of gravitational energy to rotational energy can lead to unphysical values if the initial resolution is poor, i.e., the number of initial cells N$_i<2145$ \citep{2007arXiv0709.2450C}. We are well beyond these limits. Our minimum number of cells, extending beyond the central region, is at least N$_i\geqslant16384$ cells. The Jeans length is resolved by more than N$_j>77$ cells. For even greater precision, we keep the tree structure of the top level hierarchy intact in FLASH. In other words, we initiate all simulations with one top-level block. Every simulation is continued up to twelve dynamical timescales, which is computed as

\begin{equation}
t_{\rm dyn}=\sqrt{3\pi/32G\rho}.
\label{eq:t_dyn}
\end{equation}

Where $G$ is the gravitational constant and $\rho$ is the mass density. \\

We do not include magnetohydrodynamic (MHD) effects in our models. \cite{2005JRASC..99R.132T, 2007MNRAS.382...73T} argue that these should not matter for the core formation as much as the turbulence does. However, MHD certainly is an important aspect, see \cite{2007ApJ...668.1028B, 2008arXiv0808.0986B}, that we will investigate further in the future.

\subsection{Initial Conditions}
We start the basis simulations with four metallicities $Z/Z_{\odot}=1,10^{-1},10^{-2},10^{-3}$ with respect to solar metallicity and take $Z=Z_{\odot}$ as a benchmark value. The dependence of the model outcomes on metallicity is the focus of this work.

For all simulations, we impose Gaussian turbulence with a characteristic dispersion of $\sigma=2$ km/s, comparable to observed values \citep{2001ApJ...555..178F, 2002ApJ...572..238C} and of the order of 0.3$\sqrt{GM/R}$. We do not assume the clouds to be rotationally supported but rather that turbulent motions, following the Larson laws, dominate. We adopt a Larson like scale relation for the turbulence of the form $\sigma = 2~(R/ \rm 1pc)^{0.5} ~\rm km/s$ \citep{1981MNRAS.194..809L}. \cite{2006MNRAS.365...37B} also show that molecular clouds exhibit a $\sigma \propto R^{0.5}$ velocity dispersion law. The turbulence is introduced as part of the initial conditions of the cloud and is not maintained by external forcing. The virialization of the gas that falls into local potential wells does maintain a level of super-thermal random motions for several dynamical timescales.

We run each metallicity condition with five rotational energies, i.e., $\beta_0=10^{-1},10^{-2},10^{-3},10^{-4},0$, where $\beta_0$ is the initial ratio of rotational to gravitational energy. A Keplerian rotation is initiated around the z-axis multiplied by the ratio $\beta_0$. We define

\begin{equation}
\beta_0 \equiv \frac{I\omega^2} {GM^2/r} = \frac{v^2 r} {GM}.
\label{eq:beta0}
\end{equation}

Here, $I$ is the moment of inertia, $\omega$ is the angular velocity, $v$ is the radial velocity and $M$ and $r$ are the total cloud mass and its corresponding radius. Studies show that a typical ratio for molecular cloud cores is of the order $\beta_0=0.02$ \citep{2002ApJ...572..238C}. These authors have observed 57 cloud cores and find energies within the range of $\beta_0 \approx 10^{-1}-10^{-4}$. Cosmological simulations indicate that star forming host clouds have rotation energies of the order of $\beta_0 \simeq 0.1$ \citep{2002ApJ...564...23B, 2006ApJ...652....6Y}.

We simulate an initially spherical cloud with a radius of 10 pc within a box that is periodic in both gravity and spatial boundaries. Given the larger box size of 26 pc and due to symmetric initial conditions, we do not expect nor see that the periodicity is an issue. We initialize the cloud with a mean constant density of $n\simeq10^3 \rm ~cm^{-3}$ \citep{Frieswijk2008}, since it is not likely that a well developed density profile exists before virialization. These initial conditions mean that we start with an initial cloud mass of 28.3$\times$10$^{3}$ solar masses. The initial gas temperature is put at 100 K throughout the cloud, simulating starburst regions \citep{2000ApJ...538..115S, 2005ApJ...629..767O}.

\subsection{The Simulations}
We present twenty simulations, with flat initial density profiles, where we focus on the effect of metallicity on cloud fragmentation. In a similar fashion as the work by \cite{Machida1} and \cite{2009arXiv0907.3257M}, we also investigate the importance of rotation and turbulence in these. We present 15 more simulations with the same setup to test and compare specific physical cases. \\

\noindent
{\bf Metallicity}: Four sets of five simulations are done in order to follow the collapse and fragmentation of metallicity dependent cooling functions for turbulent, rotating and non-rotating cases. These simulations have an initial cloud temperature of 100 K, which represents the high temperatures within the active environments where the molecular clouds are formed. \\
{\bf Dust and CRs in starbursts}: Five solar metallicity runs with additional heating sources; cosmic rays and dust. In massive star forming regions prominent in starbursts there can be a lot of warm dust affecting the thermal balance of the cloud. Dust can act as a cooling mechanism when the gas temperature is higher than the dust temperature. Once the temperature of the gas is lower than the dust, and this is the case for photon dominated regions (PDRs) with $A_V > 5$ mag, it will become a source of heating. \\
Cosmic rays can ionize and heat the gas, but their contribution is generally thought to be low. In starbursts, the impact of CR-heating is stronger due to the increased production by massive stars, i.e., supernovae. \\
{\bf Isothermal at 25 K}: Five runs kept the temperature nearly constant at a starting cloud temperature of 25 K. Representing an environment where the temperature is higher than what we observe in the Milky Way. For these simulations, the temperature is not allowed to go below 25 K and the compressional heating term is removed. The results are compared to cases where the thermal balance is incorporated. \\
{\bf Isothermal at 10 K}: A second series of near-isothermal runs are initiated at T=10 K in order to compare to Milky Way conditions and to check the impact of the isothermal condition. \\

\subsection{Chemistry, Cooling and Heating}
\begin{figure}[htb]
\flushleft
\includegraphics[scale=0.24]{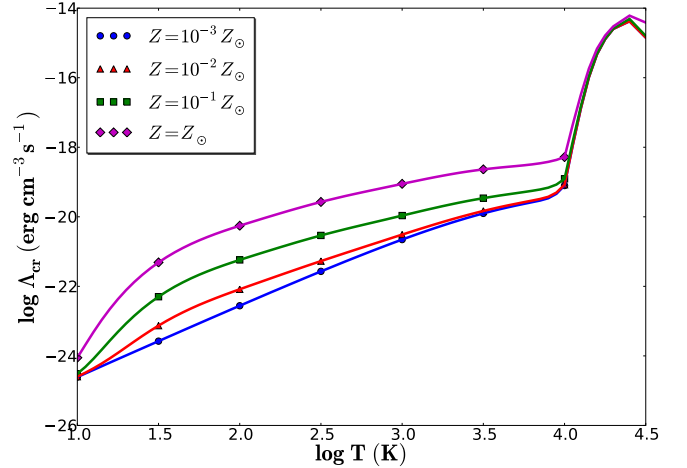}
\caption{Four cooling curves by Meijerink \& Spaans corresponding to four metallicities $Z = Z_{\odot}, 10^{-1}Z_{\odot}, 10^{-2}Z_{\odot}$ and $10^{-3}Z_{\odot}$. The cooling rates ($\Lambda_{\rm cr}$) correspond to a number density of 10$^4$ cm$^{-3}$.}
\label{fig:coolingcurve}
\end{figure}

In our simulations we use a detailed cooling function created with the \cite{2005A&A...436..397M} code. The metallicity dependent cooling includes fine-structure emissions from carbon and oxygen as well as molecular lines from species like CO and H$_2$O. The level populations have been corrected for LTE effects above 10$^{3}$ cm$^{-3}$. The chemistry includes gas phase and grain surface formation of H$_2$ and HD \citep{2004ApJ...611...40C, 2009A&A...496..365C} and line trapping due to the initial column (N$_{\rm H}\sim \rm 10^{22.5} ~cm^{-2}$ and $\sigma$ = 2 km/s) through the cloud, as described in \cite{2005A&A...436..397M} and using the multi-zone escape probability method of \cite{2005A&A...440..559P}. Line trapping during cloud collapse is ignored (see the discussion). A background radiation field has been included that conforms to 0.1 G$_{0}$ \citep[in units of the Habing field,][]{1968BAN....19..421H} in the dwarfs and 30 G$_{0}$ for the starbursts \citep{1997ApJ...483...87S, 2007MNRAS.374L..29K}, see figure \ref{fig:coolingcurve}.

At large column densities, cosmic ray heating can play an important role. For our heating by cosmic rays, we use the definition of \cite{2005A&A...436..397M} as presented in equation \ref{eq:cr}, which scales linearly with the density.

\begin{equation}
\Gamma_{CR} = 1.5 \times 10^{-11} ~\zeta $ n(H$_2$)$  $~~erg cm$^{-3} $ s$^{-1}.
\label{eq:cr}
\end{equation}

We have taken the cosmic ray ionization rate per H$_{2}$ molecule $\zeta$ as 9$\cdot$10$^{-16}$ s$^{-1}$. This corresponds to a starburst environment with a background star formation rate of the order of 30 M$_{\odot}$ per year, with the appropriate scaling for dwarf galaxies \citep{2005ApJ...626..644S}.

\begin{figure*}[htb!]
\centering
\begin{tabular}{c c}
\begin{minipage}{7.5cm}
\includegraphics[scale=0.38]{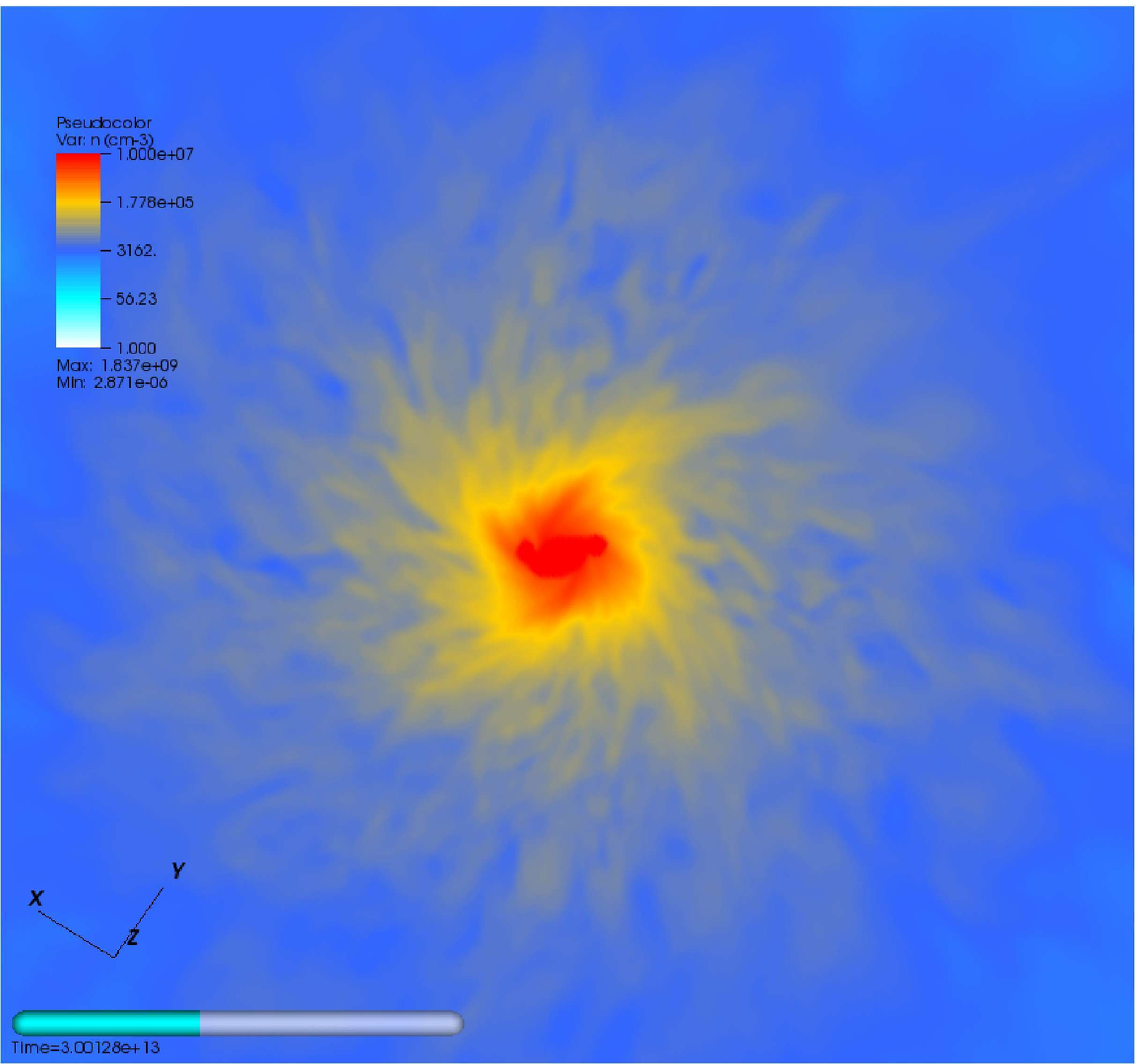}
\end{minipage} &

\begin{minipage}{8cm}
\includegraphics[scale=0.38]{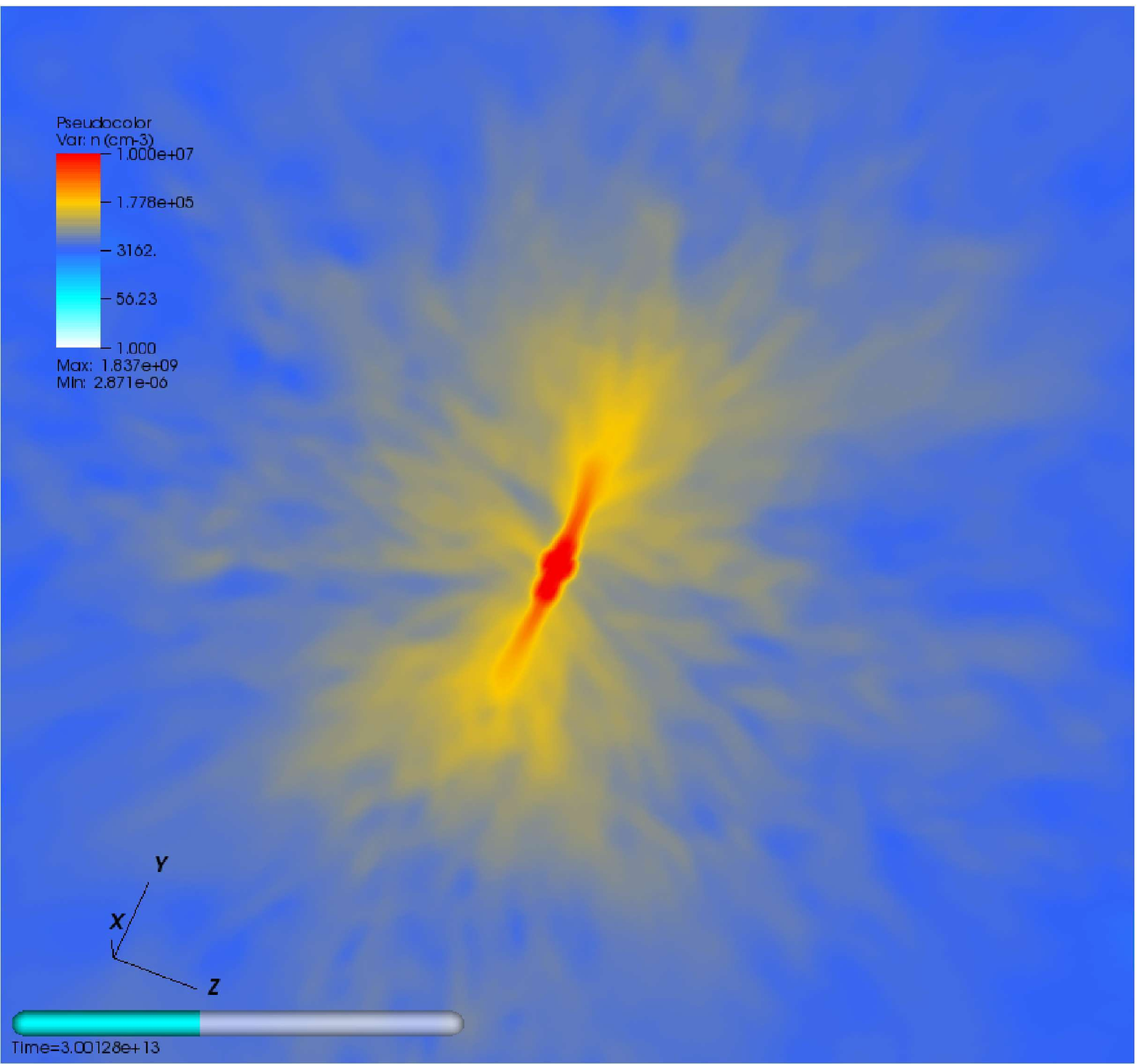}
\end{minipage} \\

\end{tabular}
\caption{A 2D slice of a 3D cloud collapse simulation at $t \simeq t_{\rm dyn}$. Plotted variable is density. Left: A face-on view of the disk. Right: An edge-on view of the disk.}
\label{fig:Disk}
\end{figure*}

Besides the heating by cosmic rays, heat can be transferred through collisions between dust and gas. The gas-dust temperature difference is important here. We add gas-grain collisional heating to our simulations as given by \cite{1989ApJ...342..306H, 1979ApJS...41..555H} and \cite{2005A&A...436..397M};

\begin{eqnarray}
\Gamma_{\rm coll.} &=& 1.2 \times 10^{-31} n^2 \left( \frac{\rm T_{\rm k}} {1000} \right)^{\frac{1}{2}} \left( \frac{100\ {\rm \AA}} {a_{\rm min}} \right)^{\frac{1}{2}} 
\\ \nonumber
& & \times \left[ 1 - 0.8\exp \left( \frac{-75} {\rm T_{\rm k}} \right) \right] (T_{\rm d} - T_{\rm k})
\rm ~~erg ~cm^{-3} ~s^{-1}.
\label{eq:gas-grain}
\end{eqnarray}


In this equation, T$_{\rm k}$ is the gas kinetic temperature, n is the average number density and $a_{\rm min}$ is the minimum grain size that is taken as 10 \AA. We keep the grain size distribution fixed \cite[MRN, ][]{1977ApJ...217..425M}. The dust temperature we use, T$_{\rm d}$ = 39 K \citep{2001A&A...375..797W}, is for a starburst region \citep{2005A&A...436..397M} and is kept constant. The heat transfer between gas and grains will act as a cooling mechanism for the gas when it has a higher temperature than the dust.

Because line trapping beyond a column density 10$^{22.5} \rm ~cm^{-2}$ is not taken into account, the temperatures presented in this paper are lower limits. Above $\sim$10$^{4.5} (Z_{\odot}/Z)$ particles per $\rm cm^{-3}$, heating by dust dominates. In this case, line trapping becomes less important because the $n^2$ dependence of the gas-dust coupling prevails over the $n$ dependence, or weaker, of line emission in LTE. The fine intricacies of cooling and heating together with gravity determine the main results of this paper.

\section{Results}
First, we perform three 3D simulations at refinement levels 9 and 10 (2048$^3$ and 4096$^3$ cells). We do these simulations to see if the extra dimension produces significant differences in the results as compared to 2D. Due to one less degree of freedom, 2D simulations can result in a higher merger rate and thus fewer fragments with higher masses in the end. The three 3D turbulent simulations were carried out for modest rotational energies, $\beta_0 = 10^{-1}$, $\beta_0 = 10^{-3}$ and one non-rotating initial condition, to avoid excessive influence of rotational motions on the outcomes. These runs are slightly lower in resolution compared to the 2D simulations and we follow them up to 4 $t_{\rm dyn}$. In the two rotating cases, we see a disk forming within a time of the order of a free fall timescale $t \approx t_{\rm dyn}$, see figure \ref{fig:Disk}. After $t_{\rm dyn}$, one effectively has a 2D system. The non-rotating simulation, however, does not exhibit significant flattening. Nevertheless, we do observe that the results, even for this case, is very similar to its 2D counterpart for solar metallicity. It has been verified that fragmentation starts on the same timescales as for the 2D simulations, $t \gtrapprox t_{\rm dyn}$. The number of fragments is somewhat higher while their masses are comparable to their 2D counterparts. We emphasize that metallicity effects have our interest and feel that 2D simulations suffice to capture any salient features.

\begin{figure*}[htb]
\centering
\includegraphics[scale=1.05]{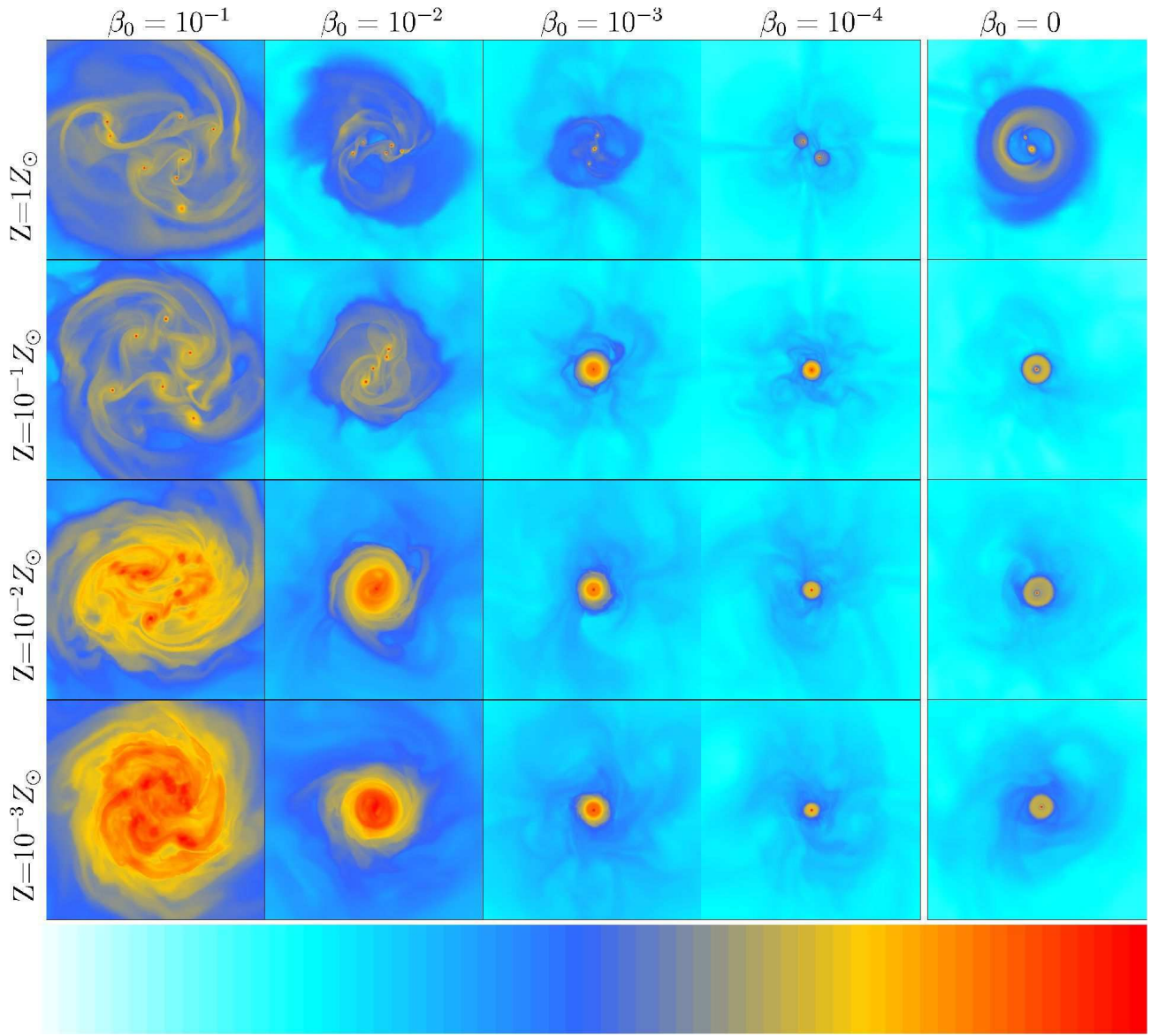}
\caption{Density plots of the final stage ($t=12t_{\rm dyn}$) of every simulation of metallicity against rotational energy. Orange to red depicts high density, typically higher than $\rm 10^{4} ~cm^{-3}$ and up to $\rm 10^{9} ~cm^{-3}$. Light blue portrays the densities smaller than $\rm 1 ~cm^{-3}$. Top to bottom: Decreasing in metallicity, $Z=Z_{\odot}, 10^{-1}Z_{\odot}, 10^{-2}Z_{\odot}, 10^{-3}Z_{\odot}$. Left to right: Decreasing in rotational energy $\beta_0=10^{-1},10^{-2},10^{-3},10^{-4},0$. 
All images have boxsizes of 4x4 parsec.}
\label{fig:basisset}
\end{figure*}

The final stage of every simulation is analyzed for the number of fragments (cores) and their masses. The results of all basis simulations in metallicity Z and initial turbulent and rotational energy, are shown as density images after 12 theoretical dynamical timescales in figure \ref{fig:basisset}.

We find the fragments by using an algorithm called \textit{isovolume} within the visualization tool \textit{VisIt} \citep{Childs:2005:ACS}. A given density threshold value lets the algorithm select the connected components above this threshold. Putting a density cap at 10$^{5}$ cm$^{-3}$ to select the dense cores \citep{2001Natur.409..159A, 2008A&A...492..703C}, we were able to count the number of fragments and determine their masses. The choice of threshold density matters for the number of fragments one derives and their masses. However, we find that there is no significant change in the number of fragments for a spread of one order of magnitude in our threshold density, while the fragment masses do decrease with increasing threshold density. Clumps and fragments can also be identified from their deep potential wells like the bound p-cores method by \cite{2009arXiv0903.3240S}. The fragment sizes range between r=6,000 AU and 84,000 AU. A complete overview of the fragments and their masses can be found in table \ref{tab:fragments}.

\begin{table*}[htb]
\begin{center}
\caption{Analysis of the fragments; the number of fragments and their masses for each case of metallicity versus rotational energy.}
\begin{tabular}{ccccc}
\hline \\
Metallicity & Rot. energy & No. of fragments & Average fragment mass & Individual fragment masses \\
$Z/Z_{\odot}$ & $\beta_{0}$ & $N_{f}$ & \textless$ M_{f} $\textgreater/$M_{\odot}$ & \textless$ M_{f} $\textgreater/$M_{\odot}$ \\
\\ \hline
\hline
$10^{0}$	& $10^{-1}$	& 8	& 2810.5	& 746.9, 1658.2, 2221.0, 3067.8, 3228.8, 3503.6, 3901.4, 4149.8\\
		& $10^{-2}$	& 5	& 5354.6	& 1818.4, 5505.1, 6238.3, 6363.6, 6847.7\\
		& $10^{-3}$	& 3	& 9363.1	& 6047.6, 7196.2, 14845.5\\
		& $10^{-4}$	& 2	& 14075.5	& 13699.2, 14451.8\\
		& $0$		& 2	& 14218.3	& 7390.8, 21045.8\\
\hline
$10^{-1}$	& $10^{-1}$	& 7	& 2619.4	& 17.8, 1754.1, 2918.7, 3082.4, 3474.2, 3517.1, 3566.8\\
		& $10^{-2}$	& 4	& 6050.4	& 3824.5, 4782.9, 5971.7, 9586.6\\
		& $10^{-3}$	& 1	& 27235.0	& No fragmentation\\
		& $10^{-4}$	& 1	& 27849.9	& No fragmentation\\
		& $0$		& 1	& 27693.8	& No fragmentation\\
\hline
$10^{-2}$	& $10^{-1}$	& 9*	& 783.0		& 53.7, 134.0, 231.0, 256.7, 273.0, 598.4, 632.8, 1374.9, 3492.5\\
		& $10^{-2}$	& 1	& 23935.9	& No fragmentation\\
		& $10^{-3}$	& 1	& 27067.5	& No fragmentation\\
		& $10^{-4}$	& 1	& 27795.2	& No fragmentation\\
		& $0$		& 1	& 28178.1	& No fragmentation\\
\hline
$10^{-3}$	& $10^{-1}$	& 9*	& 551.0		& 28.7, 167.2, 236.1, 308.2, 585.2, 711.1, 821.4, 1037.1, 1064.0\\
		& $10^{-2}$	& 1	& 24624.6	& No fragmentation\\
		& $10^{-3}$	& 1	& 27122.4	& No fragmentation\\
		& $10^{-4}$	& 1	& 27821.3	& No fragmentation\\
		& $0$		& 1	& 28225.5	& No fragmentation\\
\hline
\end{tabular}
    \label{tab:fragments}
\end{center}
\begin{footnotesize}The two cases marked by a star (*) in the table, represent the upper limits on the number of fragments.\end{footnotesize}
\end{table*}

\begin{figure*}[htb!]
\centering
\begin{tabular}{c c}
\begin{minipage}{8.5cm}
\thicklines
\includegraphics[scale=0.38]{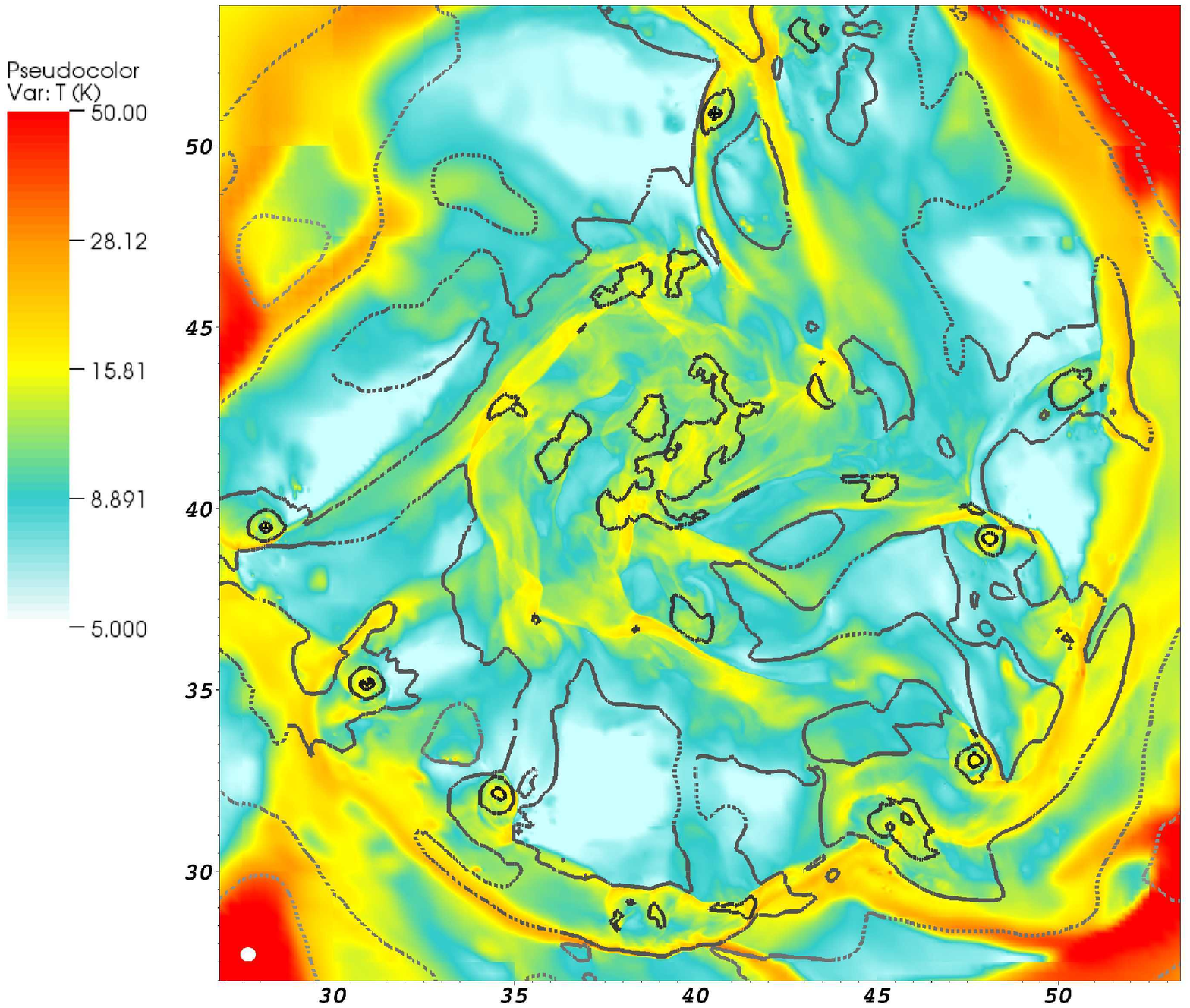}
\end{minipage} &

\begin{minipage}{8.5cm}
\includegraphics[scale=0.38]{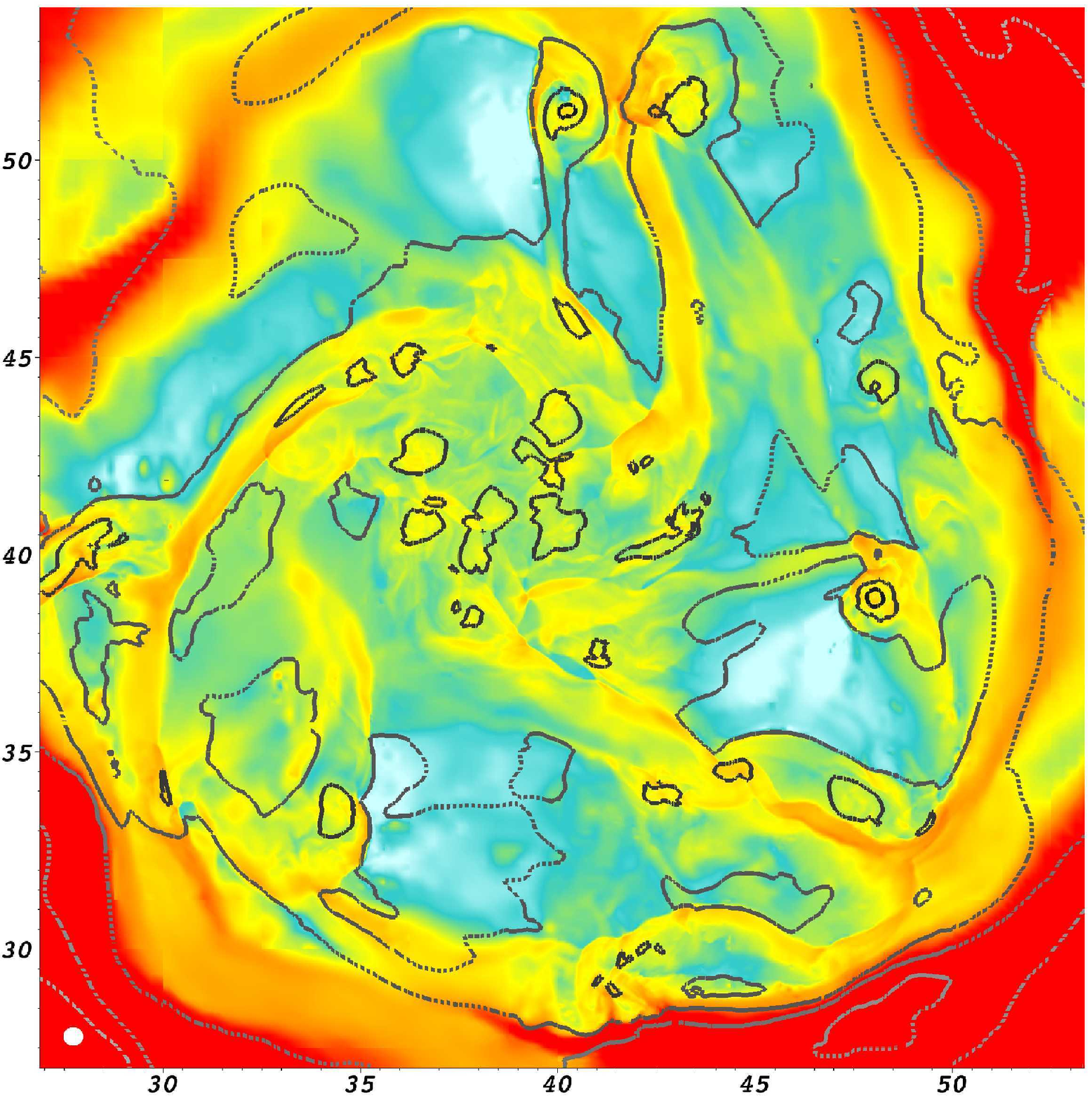}
\end{minipage} \\

\begin{minipage}{8.5cm}
\includegraphics[scale=0.38]{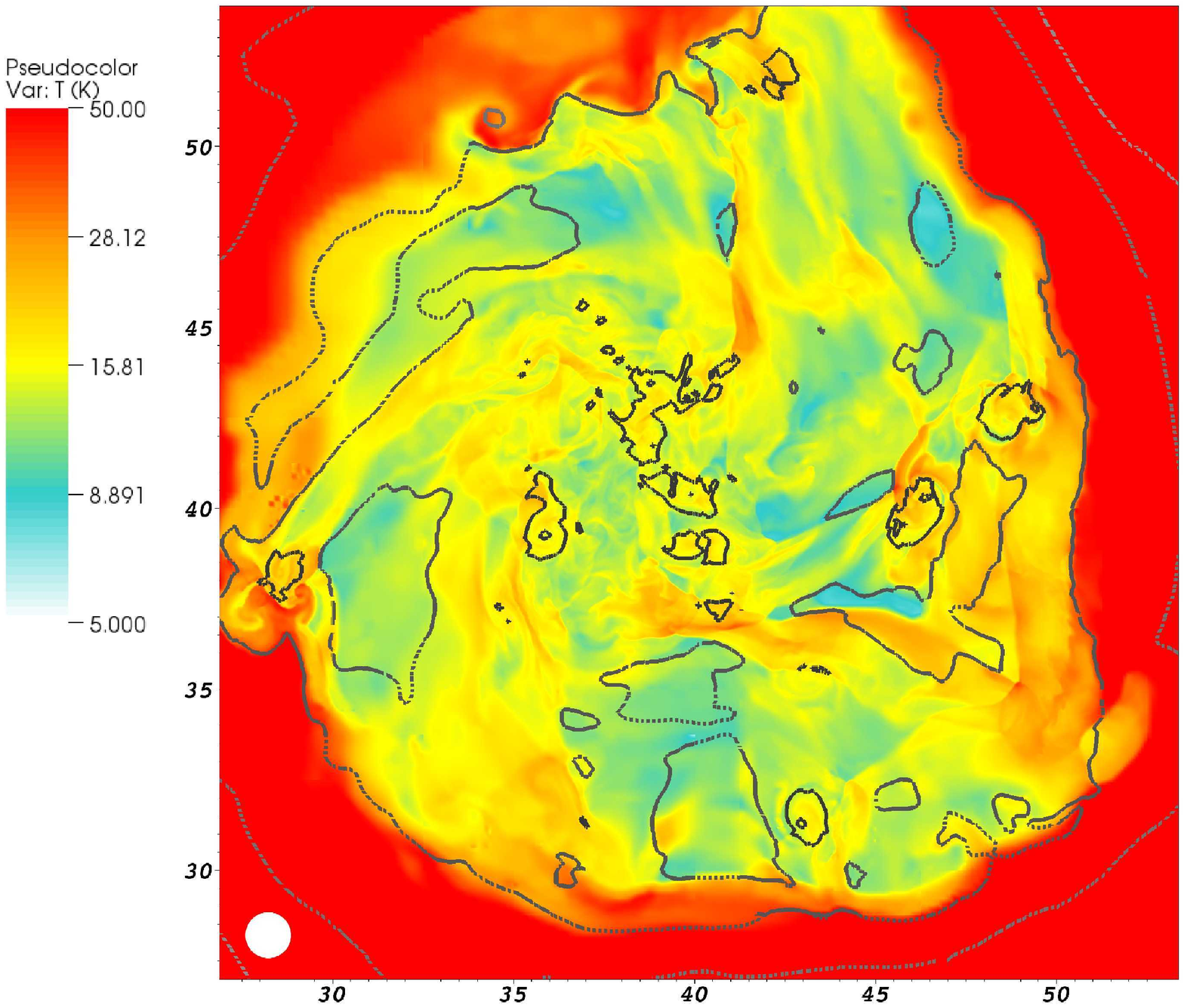}
\end{minipage} &

\begin{minipage}{8.5cm}
\includegraphics[scale=0.38]{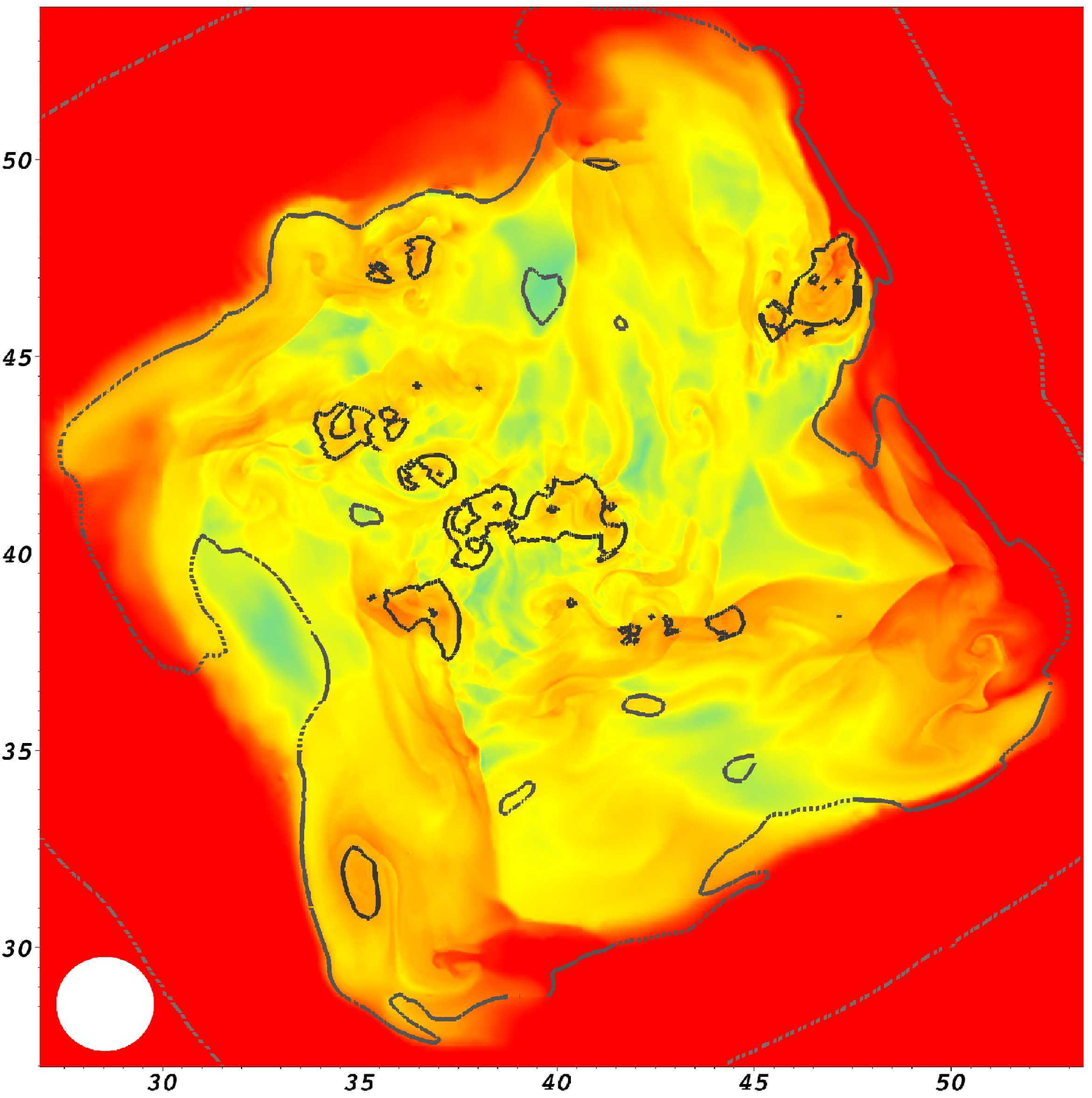}
\end{minipage} \\

\end{tabular}
\caption{Temperature plots for four different metallicities at $t=12t_{\rm dyn}$, $Z=Z_{\odot}$ (top left), $10^{-1}Z_{\odot}$ (top right), $10^{-2}Z_{\odot}$ (bottom left), $10^{-3}Z_{\odot}$ (bottom right), for $\beta_0 = 10^{-1}$. The color red represents temperatures above 50 K. Contour plots of number density are overplotted in these figures in black. Contour levels are $10^{6}, 10^{5}, 10^{4}, 10^{3}, 10^{2}, 10^{1}$ cm$^{-3}$. The axes are in units of 10$^{18} \rm ~cm$ in both directions. In the bottom left corner of each image, the Jeans length is illustrated by a white circle. To calculate the Jeans length, the average value of the temperatures and densities of the clumped regions of each image is used. These are: T=(10.0, 13.1 13.4, 19.0)[K] and $\rho$=(7.0, 3.3, 0.41, 0.22)$[\rm \times 10^{-19} g ~cm^{-3}]$.}
\label{fig:temp}
\end{figure*}

\subsection{Metallicity and Multi-Phase Structure}
We evaluate the temperature inside the cores and that of the surrounding gas of every simulation. In most cases, the cooling function is able to cool the gas efficiently. Even though there is compressional heating, the gas can cool down to temperatures as low as 10 K. The difference lies in the rate with which each model cools down or heats up. For the various metallicities, we present temperature plots in figure \ref{fig:temp}. The images are representative for every value of $\beta_{0}$. The density is overplotted with contours in these images to highlight the high density regions, and specifically, to show the location of the dense cores. The Jeans length is illustrated on the lower left side of the images\footnote{Note that the diameter of this circle corresponds to twice the Jeans length, which we refer to here as the Jeans circle. For an object to be able to fragment, it must have a size of at least twice the Jeans length and thus, must be larger than the Jeans circle.}.

We find that the temperature increases inside the fragments as the metallicity decreases. The regions where the density is higher than its surroundings have temperatures of the order of 10 K on average for $Z=Z_{\odot}$ and 20 K for $Z=10^{-3}Z_{\odot}$. We see the same trend for the lower density gas, in that the amount of cool gas, i.e. gas with temperatures lower than 50 K, decreases with Z. For a gas density above 10$^{4} \rm ~cm^{-3}$, and a gas temperature of more than 100 K, fine-structure cooling of [OI] dominates together with mid-J CO emission and water at the 20\% level. Below 100 K, low-J CO line emission is the main coolant with contributions from water on the 10\% level. Diffuse gas below 10$^{4} \rm ~cm^{-3}$ is mainly cooled by [CII].

\begin{figure*}[htb!]
\begin{tabular}{c c}
\centering
\begin{minipage}{8.5cm}
\includegraphics[scale=0.29]{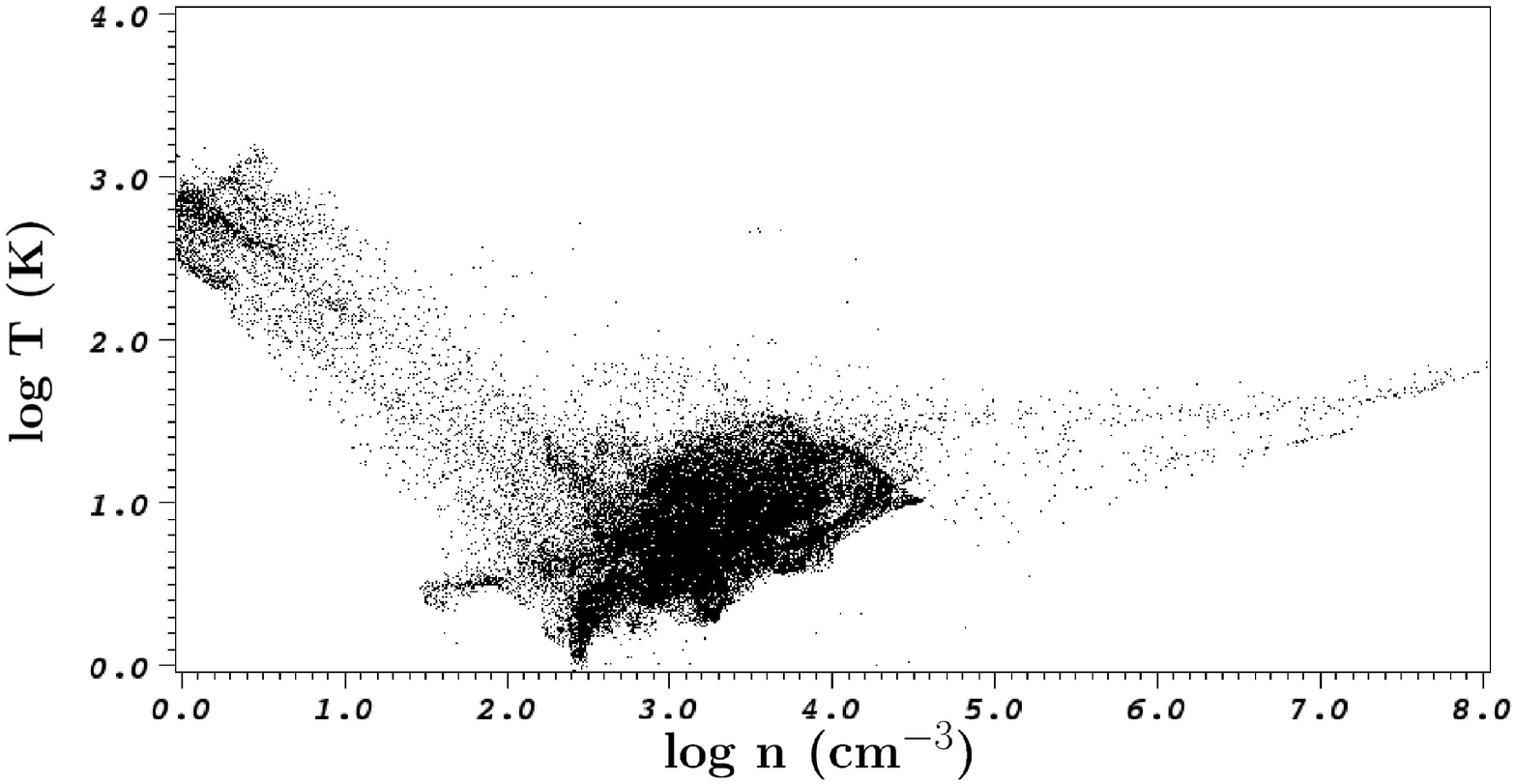}
\end{minipage} &

\begin{minipage}{8.5cm}
\includegraphics[scale=0.29]{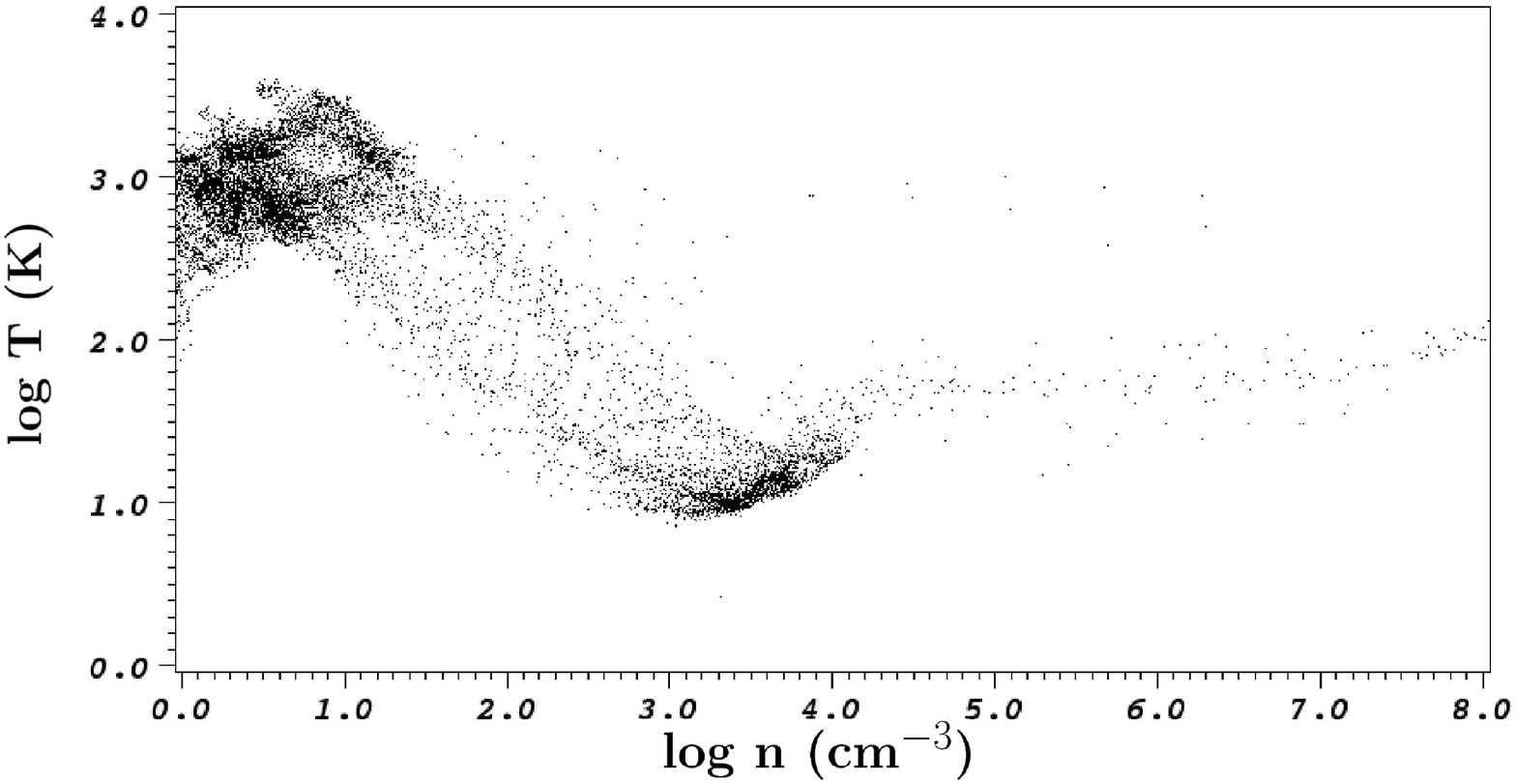}
\end{minipage} \\

\\~\\~\\

\begin{minipage}{8.5cm}
\includegraphics[scale=0.29]{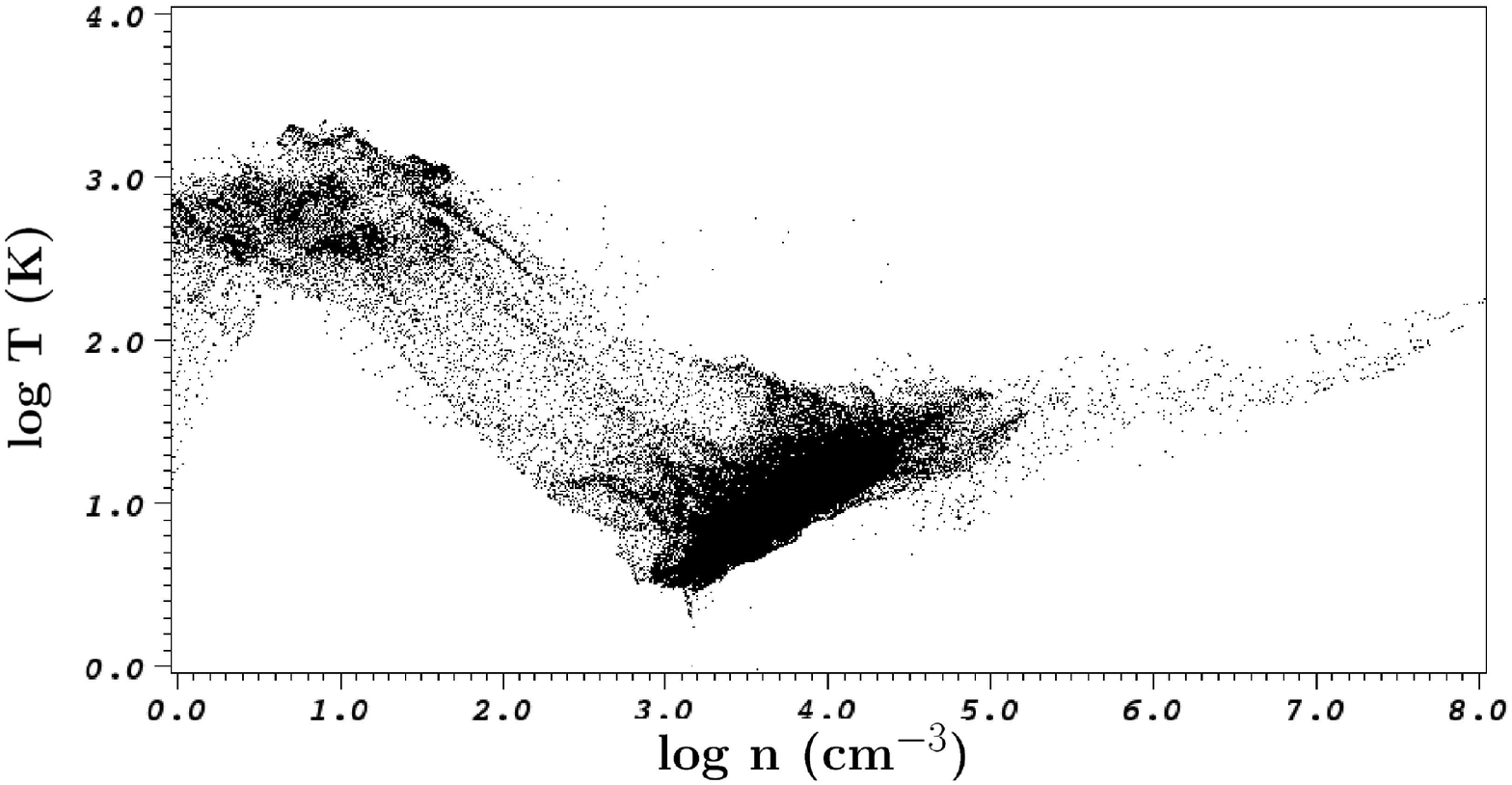}
\end{minipage} &

\begin{minipage}{8.5cm}
\includegraphics[scale=0.29]{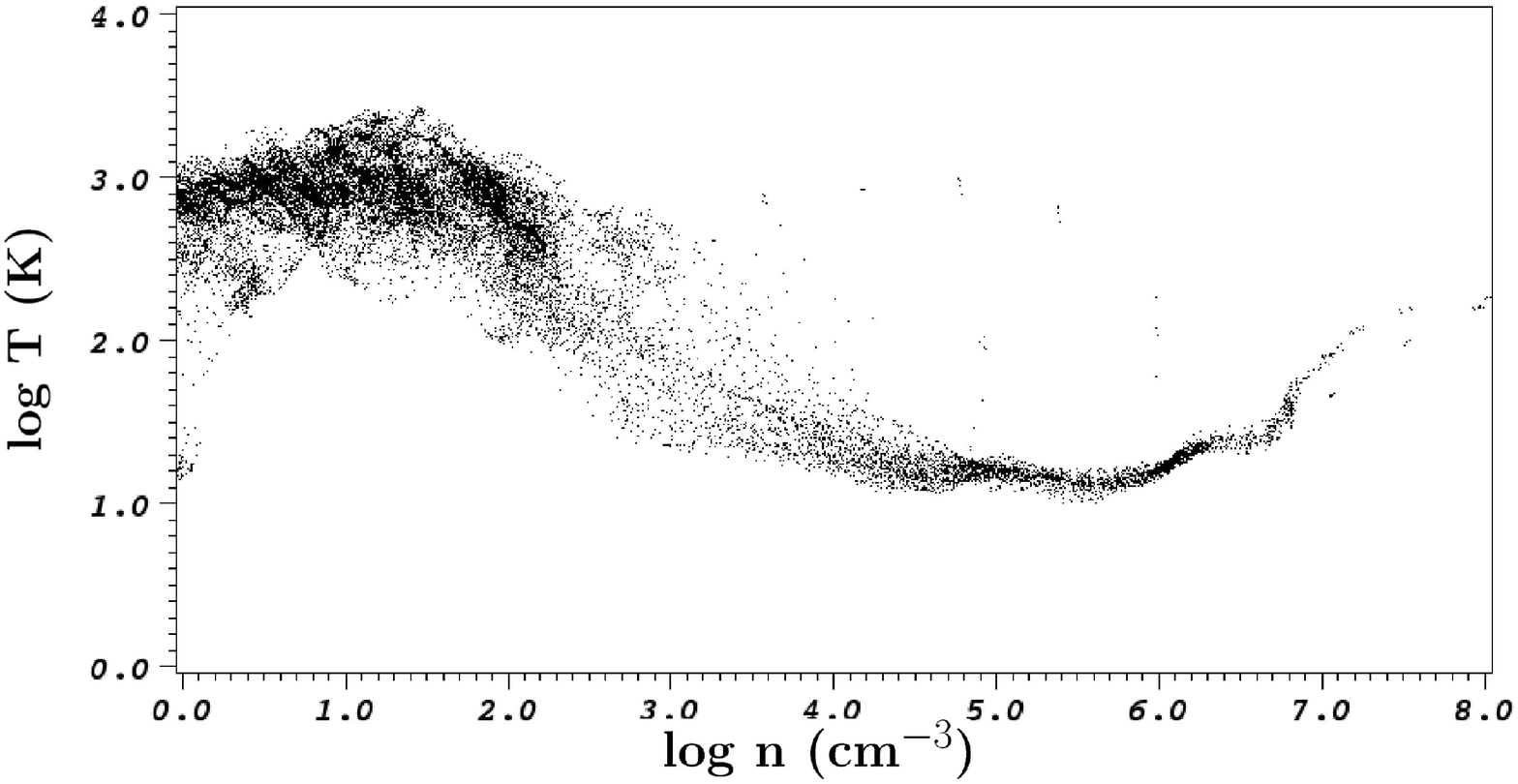}
\end{minipage} \\

\\~\\~\\

\begin{minipage}{8.5cm}
\includegraphics[scale=0.29]{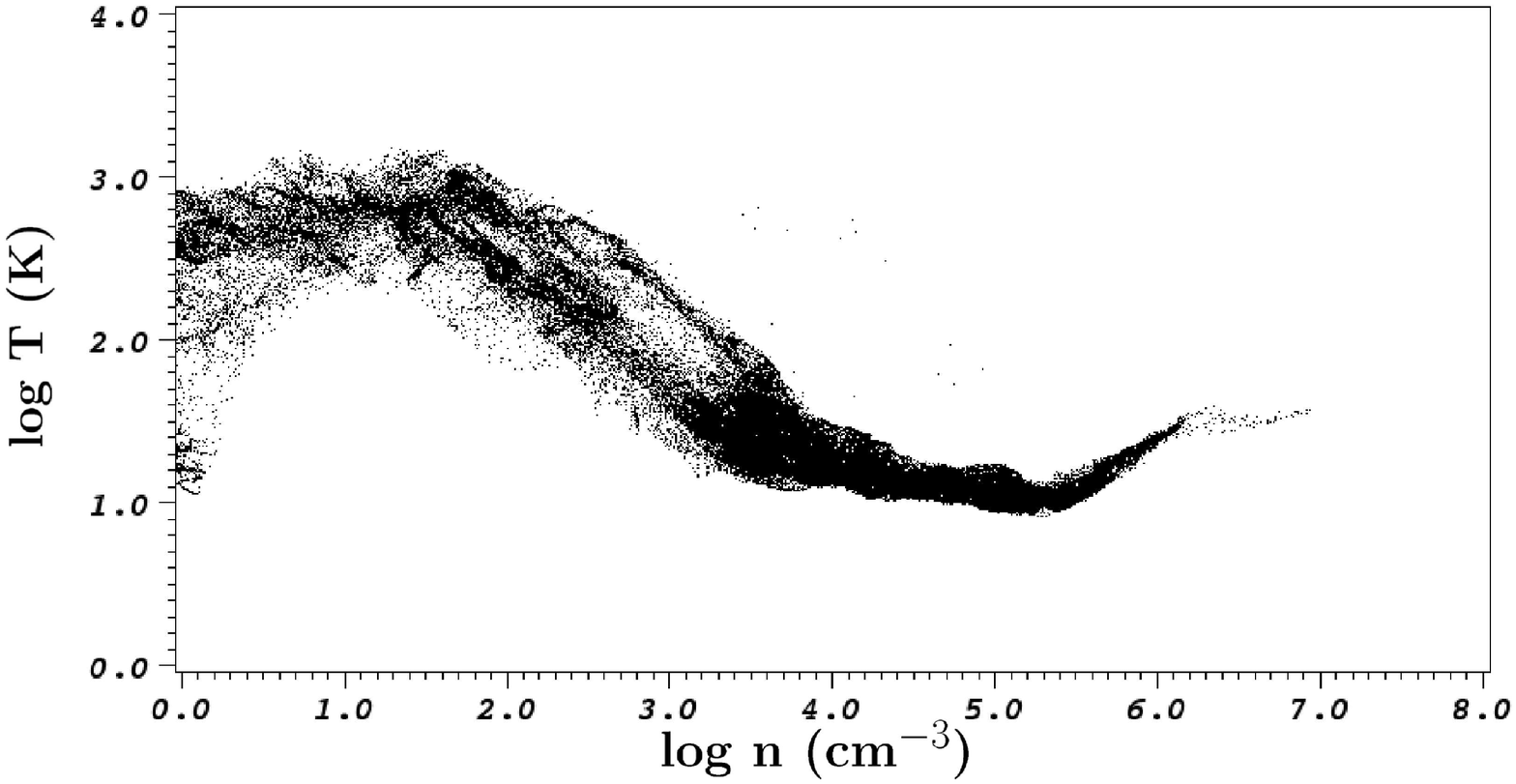}
\end{minipage} &

\begin{minipage}{8.5cm}
\includegraphics[scale=0.29]{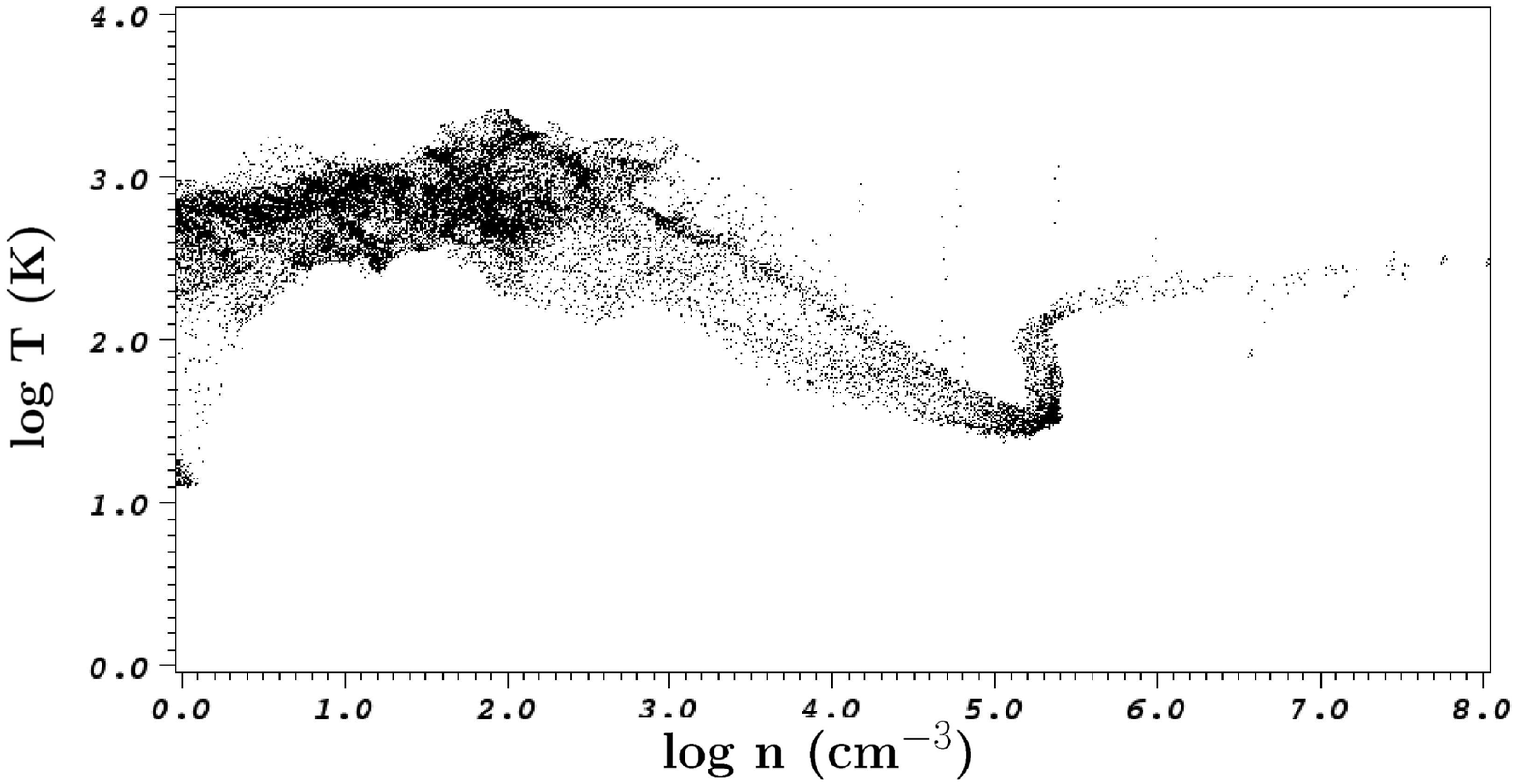}
\end{minipage} \\

\\~\\~\\

\begin{minipage}{8.5cm}
\includegraphics[scale=0.29]{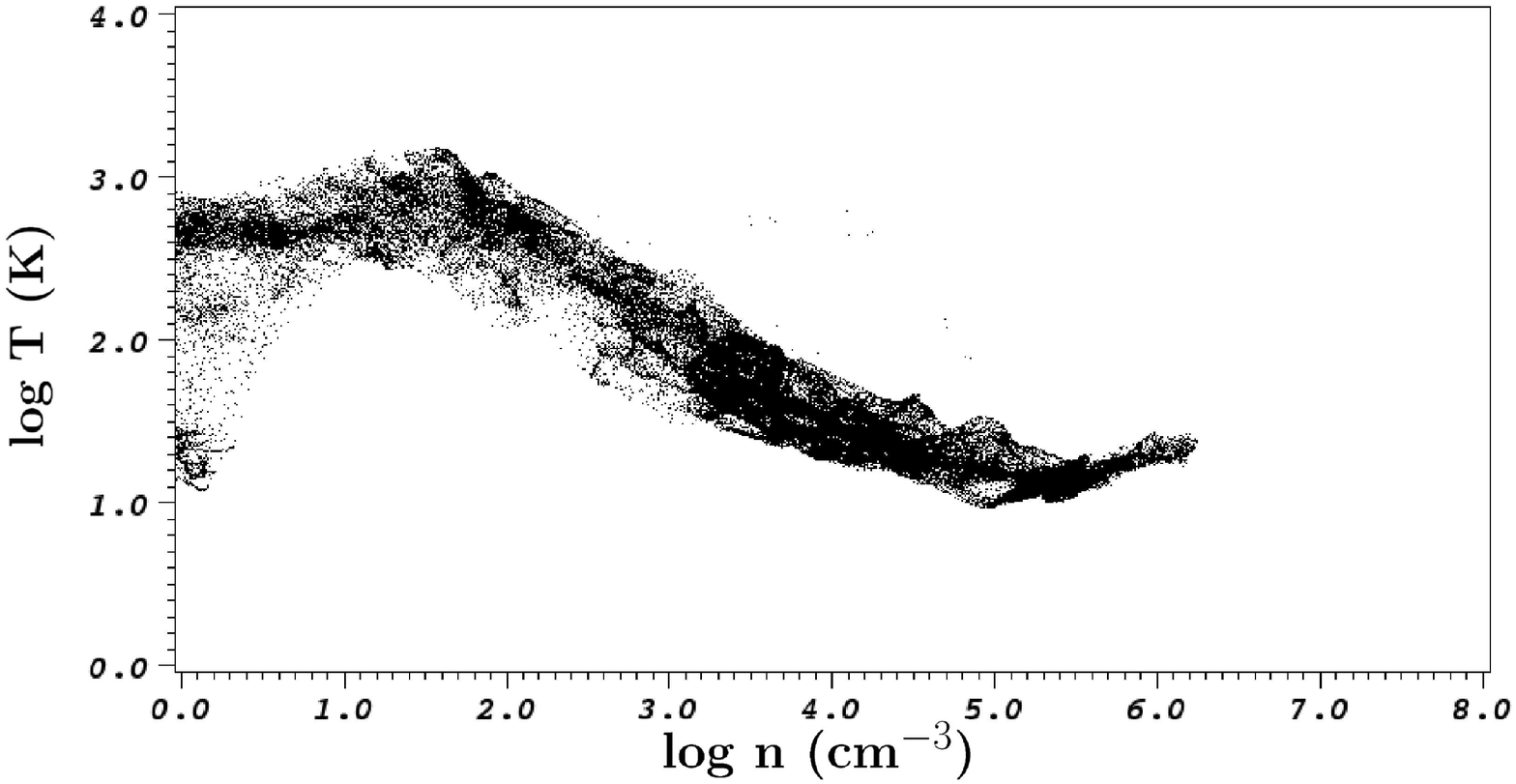}
\end{minipage} &

\begin{minipage}{8.5cm}
\includegraphics[scale=0.29]{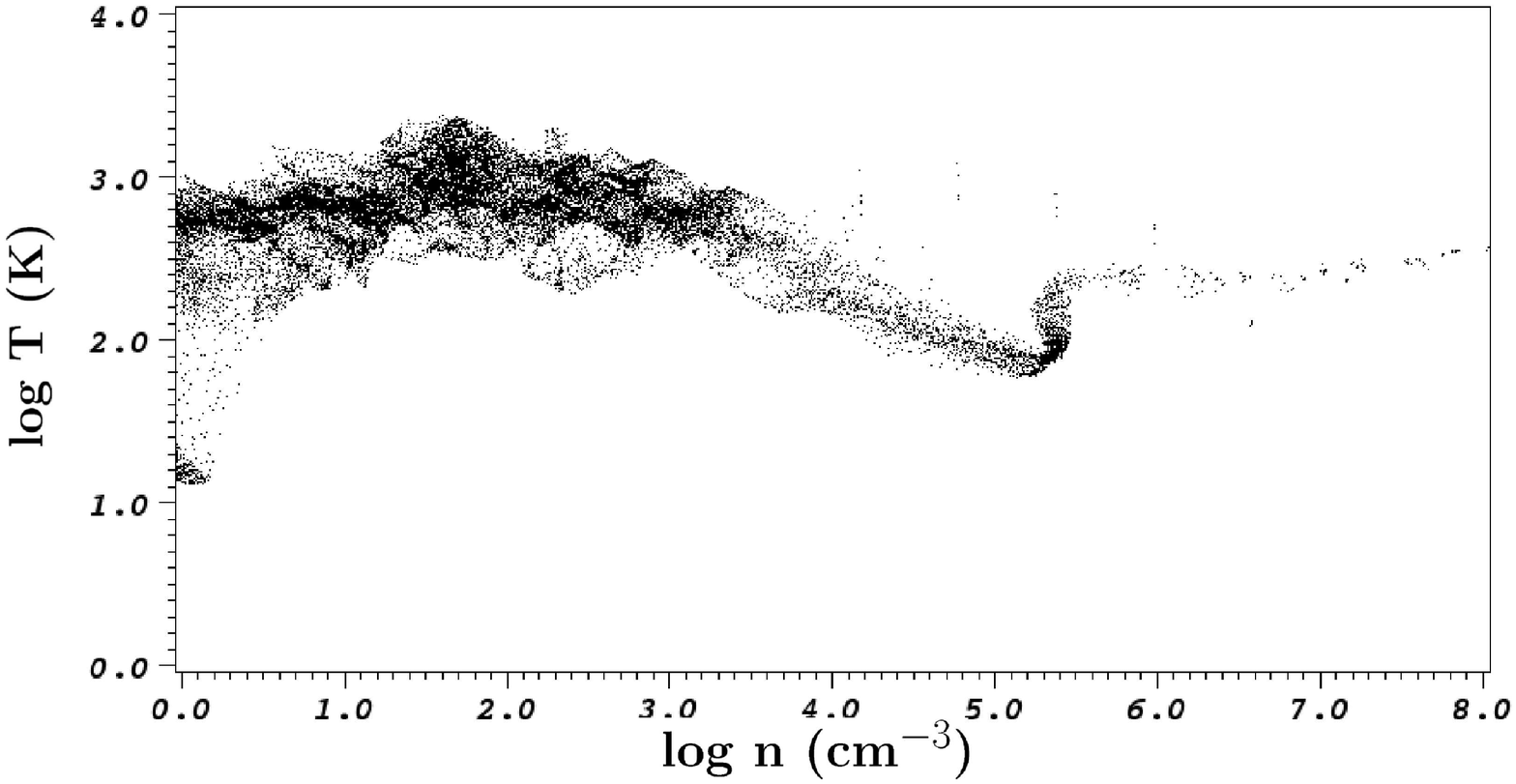}
\end{minipage} \\

\end{tabular}
\caption{The phase diagrams of each metallicity after twelve dynamical timescales for two $\beta_0$ ratios, 10$^{-1}$ (left) and 10$^{-3}$ (right). Top to bottom, highest ($Z=Z_{\odot}$) to lowest ($Z=10^{-3}Z_{\odot}$) metallicity.}
\label{fig:phasediagrams}
\end{figure*}

\begin{figure*}[htb!]
\begin{tabular}{c c}
\centering
\begin{minipage}{8.5cm}
\includegraphics[scale=0.29]{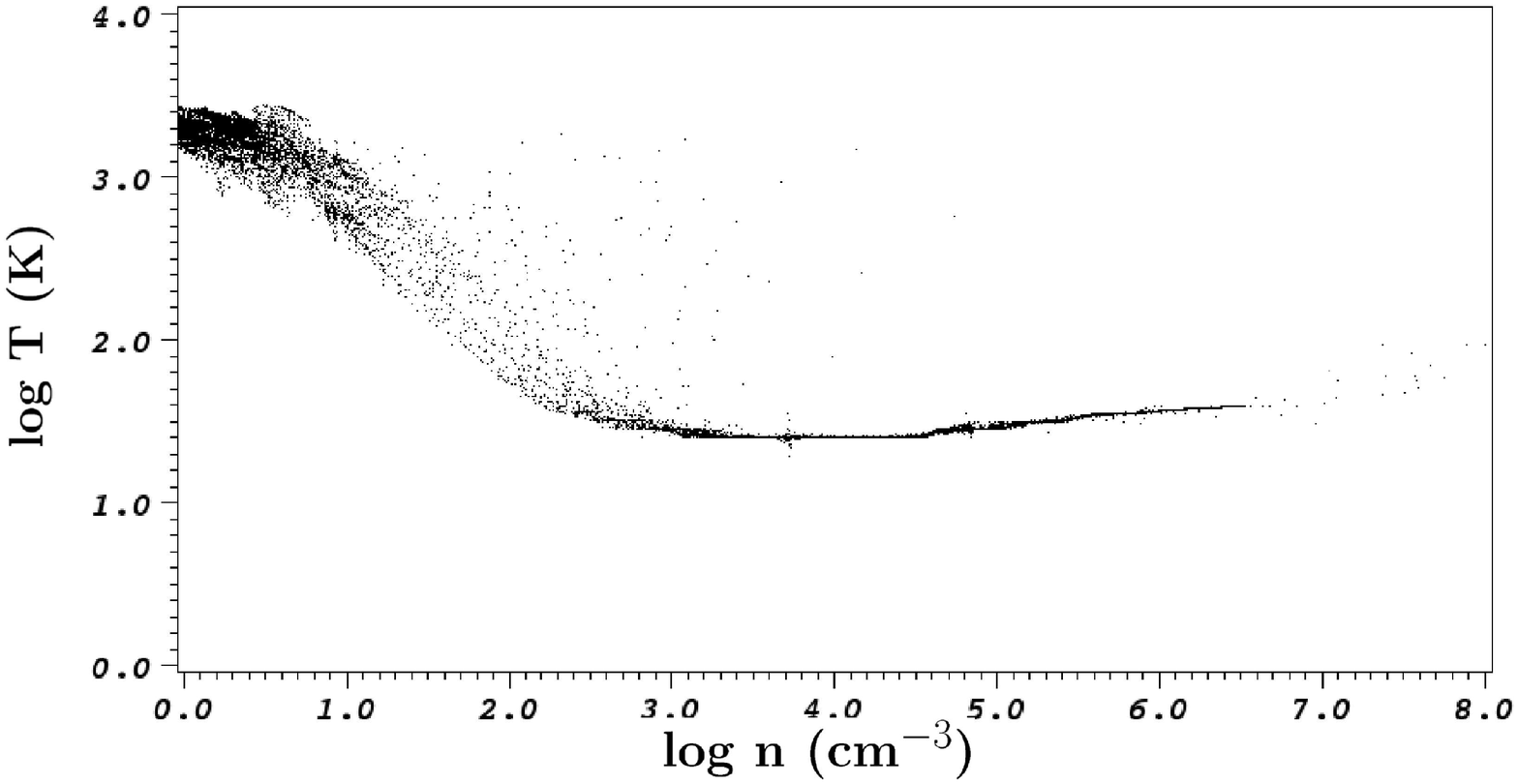}
\end{minipage} &

\begin{minipage}{8.5cm}
\includegraphics[scale=0.29]{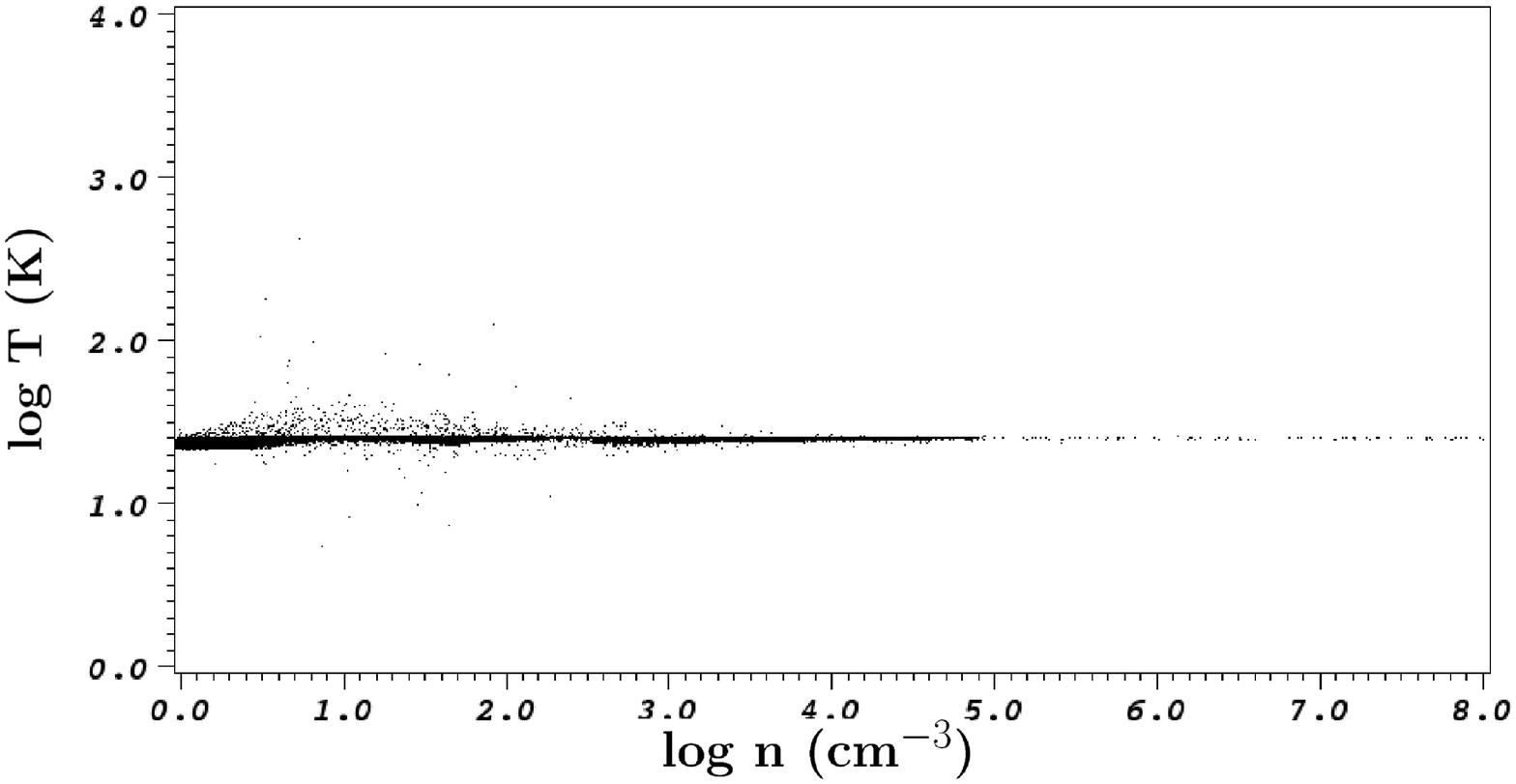}
\end{minipage} \\

\\~\\~\\

\begin{minipage}{8.5cm}
\includegraphics[scale=0.29]{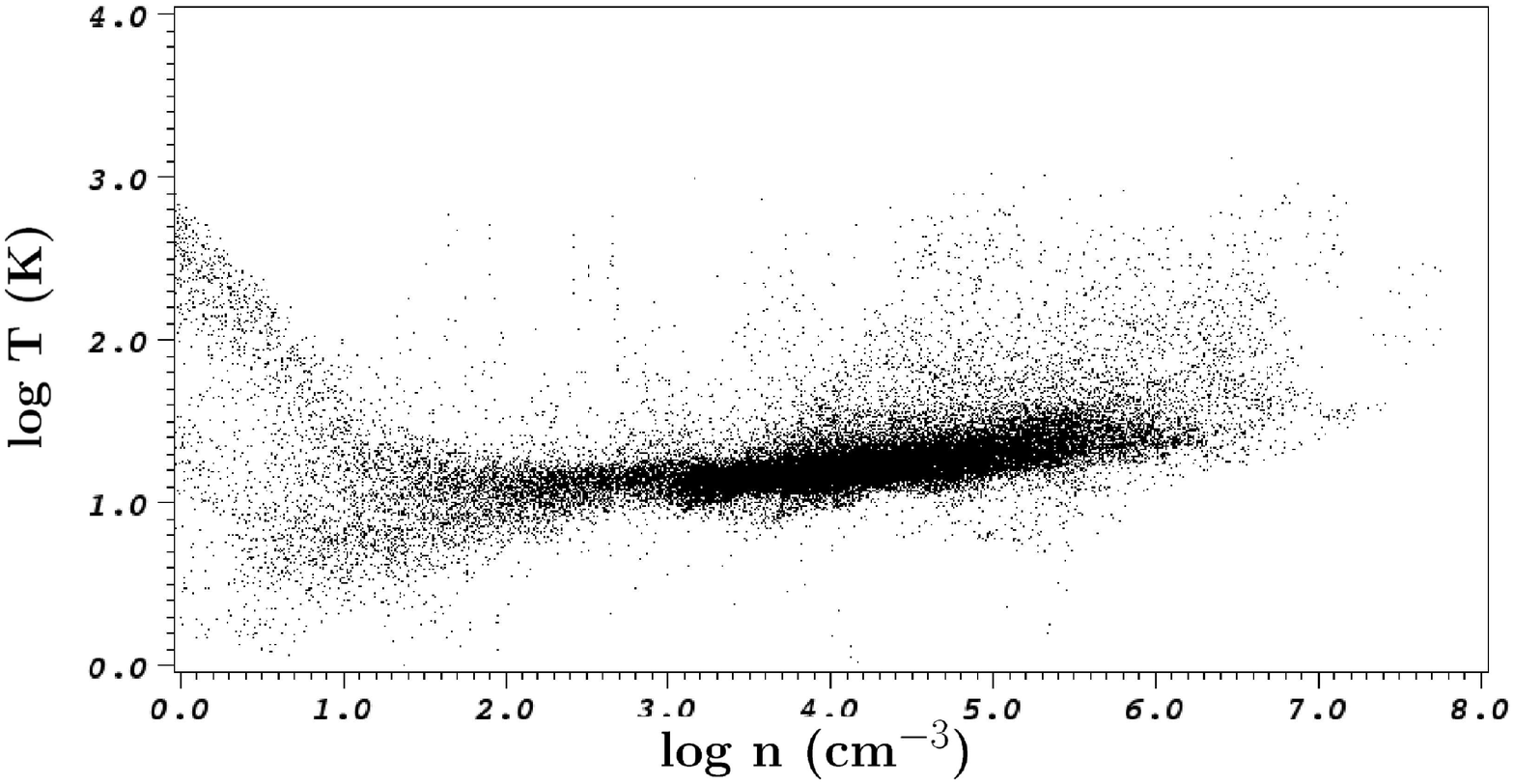}
\end{minipage} &

\begin{minipage}{8.5cm}
\includegraphics[scale=0.29]{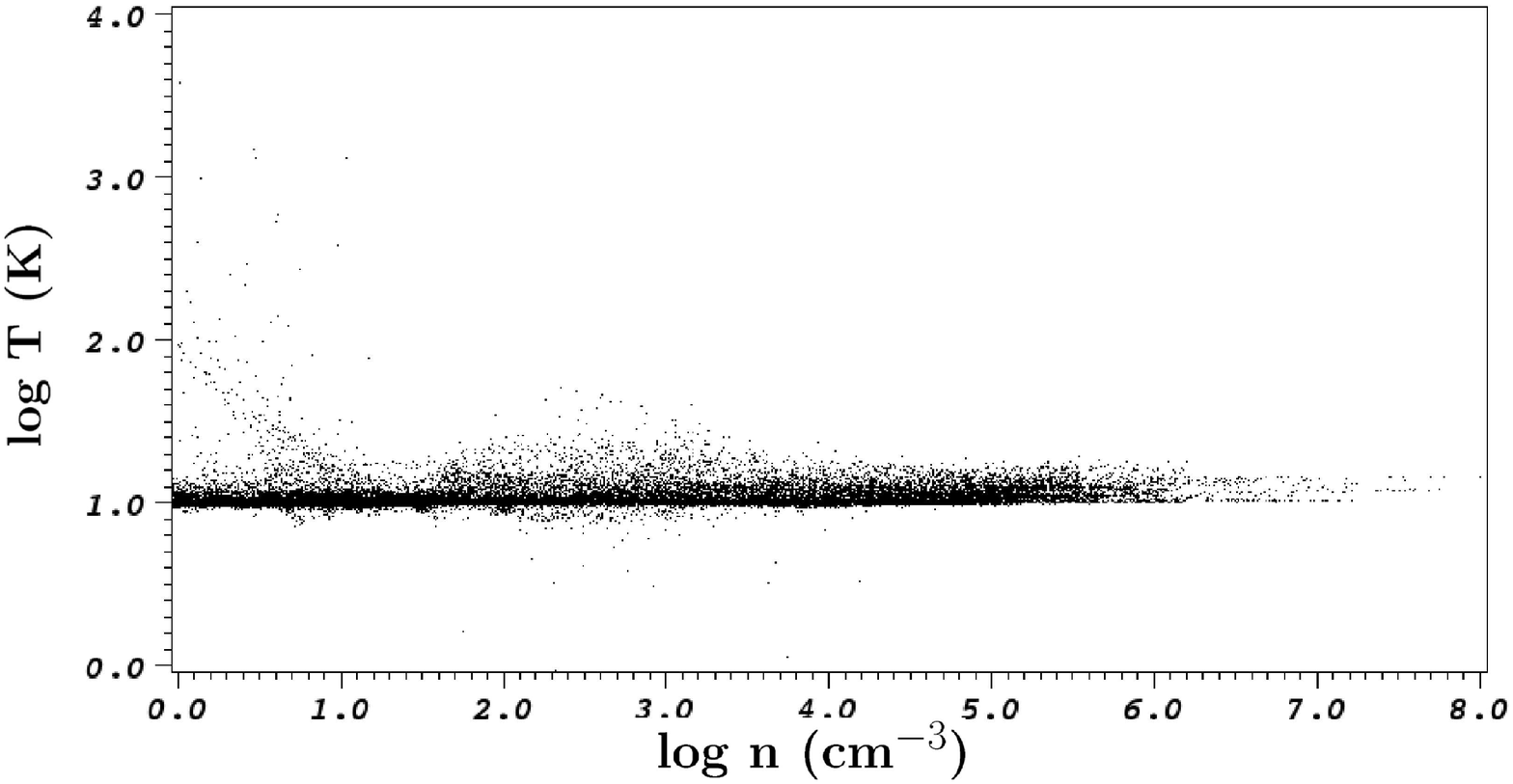}
\end{minipage} \\

\end{tabular}
\caption{The phase diagrams of several non-rotating simulations with additional physics for different environments after twelve dynamical timescales. The two figures on the right represent isothermal simulations at 25 K (top) and 10 K (bottom). The top left one is a phase diagram of the solar metallicity simulation with dust and CR heating. The bottom left figure is for a non-rotating 3D simulation also with Z=Z$_{\odot}$.}
\label{fig:phasediag-suppl}
\end{figure*}

In figures \ref{fig:phasediagrams} and \ref{fig:phasediag-suppl} we present phase diagrams, gas temperature versus density, that highlight the response of the gas to metallicity driven cooling and additional heating processes. It is immediately obvious that a cold dense phase is supported, but also that more warm gas exists at lower metallicity or higher dust temperatures. In figure \ref{fig:phasediagrams}, one immediately notices that for high metallicities, the slope $dlog(T)/dlog(n)$, is steeper between ~1 and ~10$^{3}$-10$^{4} \rm cm^{-3}$ and that the cooling time is shorter. We also note that the maximum density achieved is smaller at lower metallicities. The diagrams show that a two-phase medium exists with preferred densities of around 1 and 10$^{4} \rm cm^{-3}$. This is independent of the specific grid structure.

Above 10$^{4}$-10$^{5} \rm cm^{-3}$, compressional heating starts to dominate the thermal balance and raises the temperature. The increase due to adiabatic heating is weakened for higher metallicities, while for lower metallicities the temperature can rise up to 1000 K, if collapse is not halted completely. For low turbulent/rotational energies and in the case of zero rotation, there is a distinct curvature at a single density present. This is where the temperature rises quickly due to compression so that the inefficient cooling cannot compensate for this fast increase. At this point, the object is thermally supported until it reaches higher temperatures where the cooling gets increasingly more effective.

There are always accretion shocks with high temperatures in their wake. This is important for the clouds evolution and fragmentation. We find shocks in low density gas and occasionally near the fragments. They consist of thin layers of slightly overdense gas with high temperature and a steep velocity gradient. Shock waves are more prominent when the collapse is rapid. Shock heating can dominate under such conditions.

\subsection{The Effect of Metallicity, Turbulence and Rotation on Fragmentation}
We notice the immediate result that higher metallicity and higher rotational energy leads to more fragmentation, also found by \cite{2008ApJ...672..757C}. A similar in-depth work on this has also been performed by \cite{Machida1} and \cite{2009arXiv0907.3257M}, at higher densities. \\

\begin{figure*}[htb!]
\centering
\begin{tabular}{c c}
\begin{minipage}{8.5cm}
\includegraphics[scale=0.22]{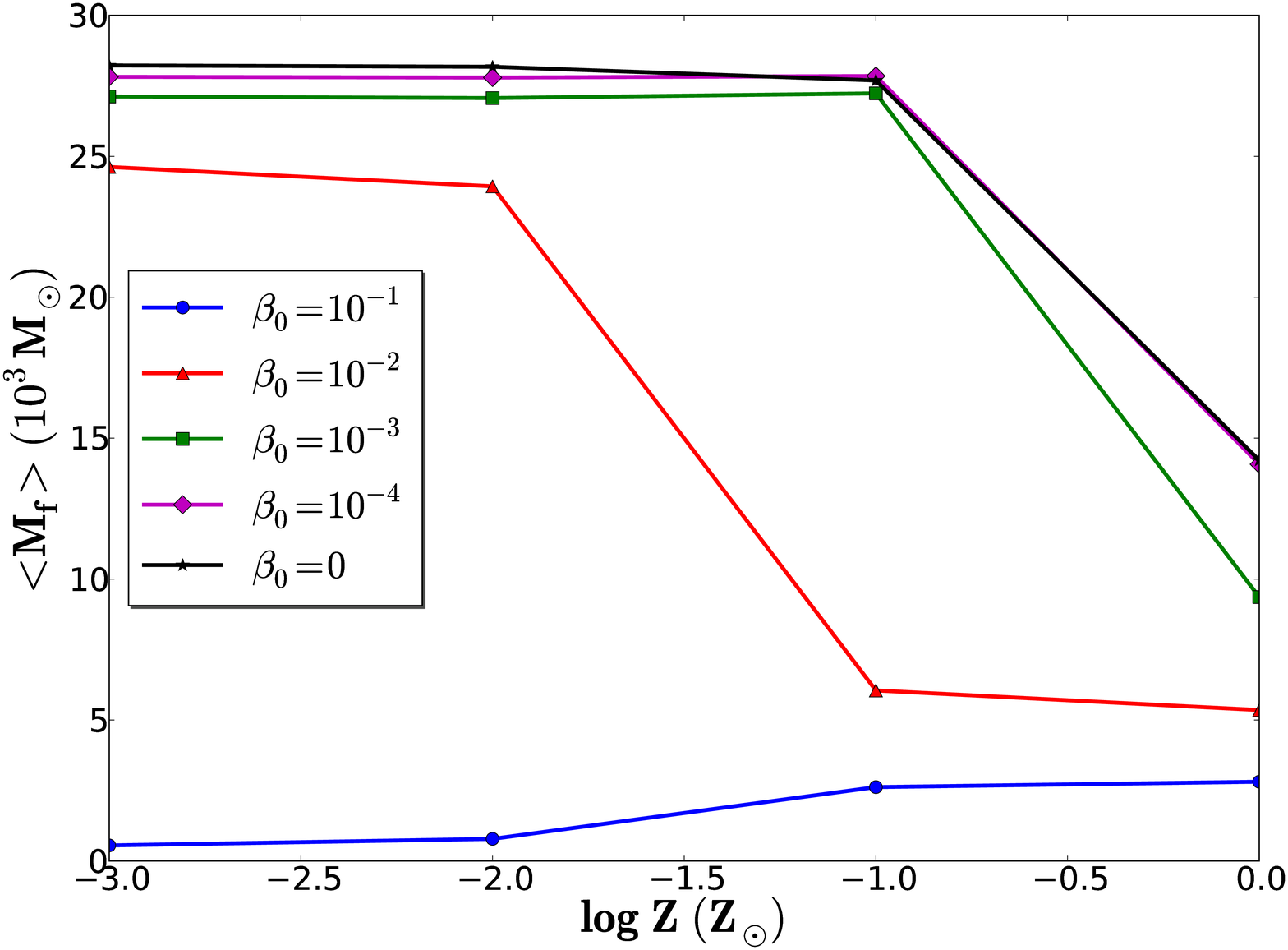}
\end{minipage} &

\begin{minipage}{8.5cm}
\includegraphics[scale=0.22]{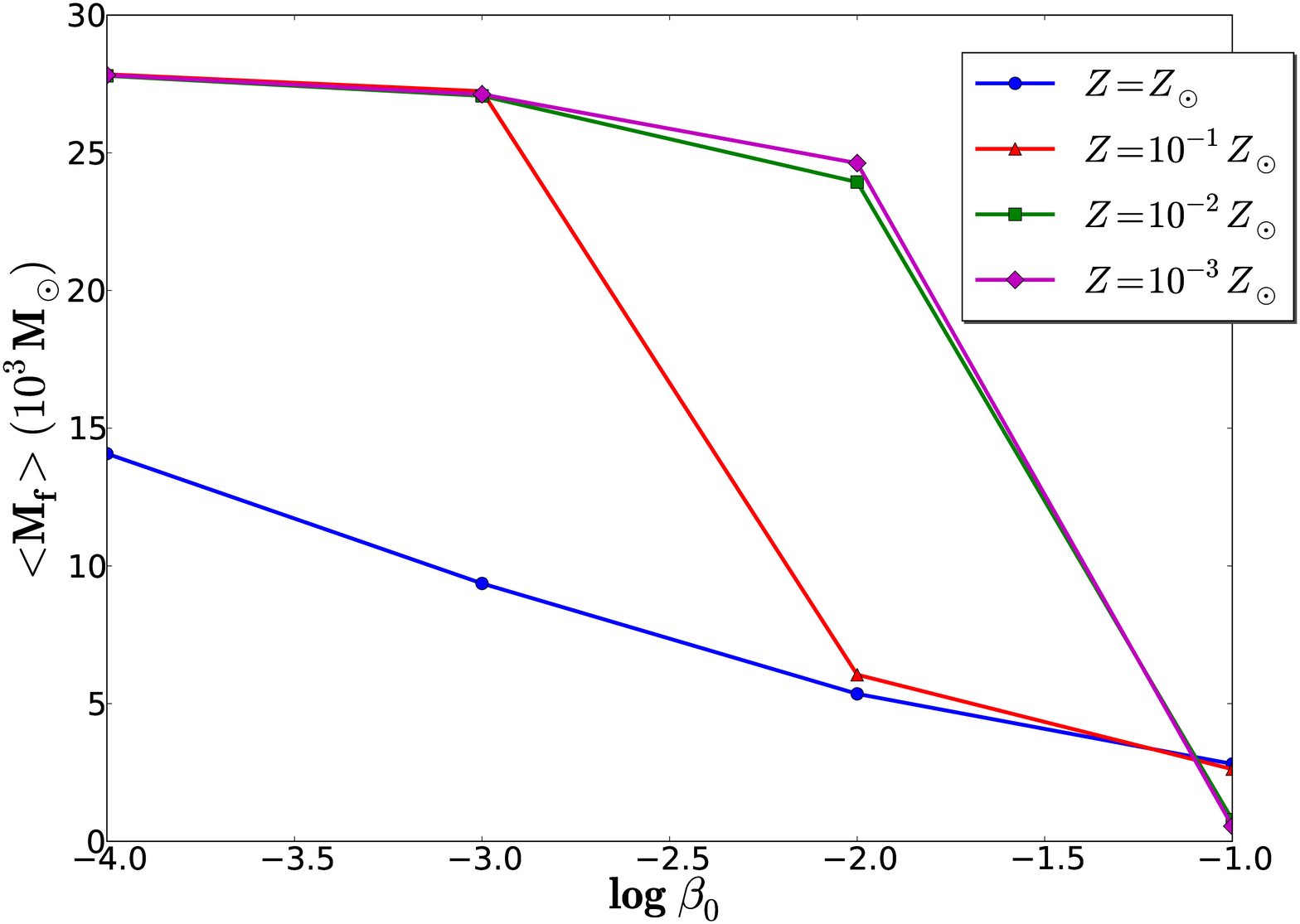}
\end{minipage} \\

\end{tabular}
\caption{Average fragment mass (\textless$ M_{f} $\textgreater) with initial conditions as determined after twelve dynamical times. Left: \textless$ M_{f} $\textgreater ~versus metallicity for each initial rotational energy. Right: \textless$ M_{f} $\textgreater ~versus rotational energy for every metallicity.}
\label{fig:efficiency}
\end{figure*}

\textit{From a metallicity perspective}, a clear trend is visible that at higher metallicities, there are more fragments and the average masses of the fragments are generally lower. The last statement is illustrated in figure \ref{fig:efficiency}. The total fragmented mass, if we sum all the fragment masses, increases with the amount of fragmentation, thus with metallicity. This is caused by the fact that the fragments accrete mass over time and more fragments accrete more mass. The compressibility of the gas also depends on the metallicity. One may understand this using a polytropic equation of state of the form $P\propto\rho^{\gamma}$, for an ideal gas $P\propto\rho T$. Increased metallicity then causes $\gamma$ (equation \ref{eq:eos2}) to be softened to unity or less since cooling is more efficient,

\begin{equation}
\gamma = 1 + \frac{dlog(T)} {dlog(\rho)}.
\label{eq:eos2}
\end{equation}

In a few low metallicity cases, not discussed in this paper, we have not found fragments with densities above the threshold criterion. In the same way, the inability to fragment holds for all low metallicity runs. While turbulence initially causes density perturbations and structures with marginally high density, the cloud lacks any development into more compact objects. We see that these undeveloped structures are rather diffuse. The internal pressure is too high for them to collapse on their own and the cooling is not efficient enough to bring down their temperatures. The latter effect causes the thermal Jeans mass, given in equation \ref{eq:jeans}, to become very high for these cases and prevents further collapse,

\begin{equation}
M_J = \left( \frac{3} {4\pi} \right)^{\frac{1}{2}}  \left( \frac{5k} {G\mu m_h} \right)^{\frac{3}{2}}  \frac{T^{\frac{3}{2}}} {\rho^{\frac{1}{2}}}.
\label{eq:jeans}
\end{equation}

\noindent
In solar mass units, the Jeans equation can be formulated as \citep{Frieswijk2008};

\begin{equation}
M_J \simeq 7.5 (\frac {T} {10 K}) (\frac {n_{H_2}} {10^4 cm^{-3}})^{-\frac {1} {2}} ~(M_{\odot}).
\label{eq:jeans2}
\end{equation}

\begin{figure}[htb]
\includegraphics[scale=0.5]{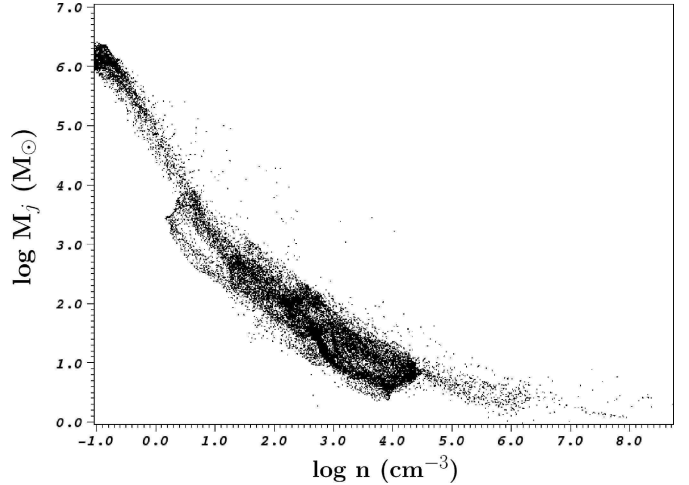}
\caption{A diagram that shows the Jeans mass as a function of number density for a non-rotating solar metallicity run at $t=12t_{\rm dyn}$.}
\label{fig:jeans}
\end{figure}

Typical Jeans masses are of the order of $\sim$10 solar masses for a 10-20 K object with a number density of $\sim$10$^{5} \rm ~cm^{-3}$. For low metallicities, where the temperature can rise up to a few 100 K, the Jeans mass lies in the range $\rm 10^{2}-10^{3} M_{\odot}$. Nonetheless, the Jeans mass can become as low as a few solar masses at higher densities despite the increase in gas temperature. A typical Jeans mass diagram of a solar metallicity simulation is shown in figure \ref{fig:jeans}. \\

\textit{From a rotational perspective}, we see that faster rotation results in more fragments in general. We find fragmentation happening for all given initial rotational energies for the highest metallicity case, due to the imposed turbulence. Even at the lowest rotational energy, collapse to a binary still occurs. Since the turbulence imparts significant angular momentum, the $\beta=0$ case looks similar to the low $\beta$ cases, except that there is more diffuse material surrounding the cores. All simulations are supersonically turbulent over the first 1-2 dynamical timescales and level of to 0.4-2 km/s after 3 $t_{\rm dyn}$. Typically, the effective line widths ($\Delta v$) at 12$t_{\rm dyn}$ are of the order of 1 km/s on the scale of the fragment ($\sim$0.1 pc). Higher metallicity tends to lead to more turbulent motions. For the lower metallicity cases we do not see fragmentation below $\omega_0 t_{\rm dyn} < 0.03$. Here, $\omega_0$ is the angular velocity that is obtained by just putting $v = \omega_0 r$ in equation \ref{eq:beta0}. Combining with equation \ref{eq:t_dyn}, we get equations \ref{eq:omega} \citep{2003ApJ...595..913M} and \ref{eq:wt}. In fact, in our studies fragmentation only occurs when the criterion of equation \ref{eq:SH1} is met. This is found by fitting a function through the points where we still see fragmentation;

\begin{equation}
\omega_0 = \sqrt{\frac{\beta_0 G M} {r^3}}
\label{eq:omega}
\end{equation}

\begin{equation}
\omega_0 t_{\rm dyn} = \sqrt{\frac{\pi^2 \beta_0} {8}}
\label{eq:wt}
\end{equation}

\begin{equation}
\omega_0 t_{\rm dyn} ~\rm exp \left( \frac{-(log(Z/Z_{\odot})-1.5)^2} {1.74} \right) \sqrt{Z/Z_{\odot}}  \approx 1.28\times10^{-2}.
\label{eq:SH1}
\end{equation}

Figure \ref{fig:efficiency}, right, shows that the average fragment mass \textless$ M_{f} $\textgreater ~is higher at lower initial rotational energies. We see here that the cloud is fragmenting and the average fragment mass is greatly reduced for lower rotational energies when the cloud has higher metallicity and is turbulent.

The specific angular momenta, $j$, of the fragments with size scales of 0.03-0.1 pc are typically a few times 10$^{-21} \rm ~cm^{2}/s$. For larger fragments ($\sim$0.3 pc) $j$ is somewhat higher and of the order of $\sim$10$^{22} \rm ~cm^{2}/s$ in agreement with \cite{1995ARA&A..33..199B}.

\subsection{The Effect of the Environment on Fragmentation}
\begin{figure*}[htb]
\centering
\includegraphics[scale=1.05]{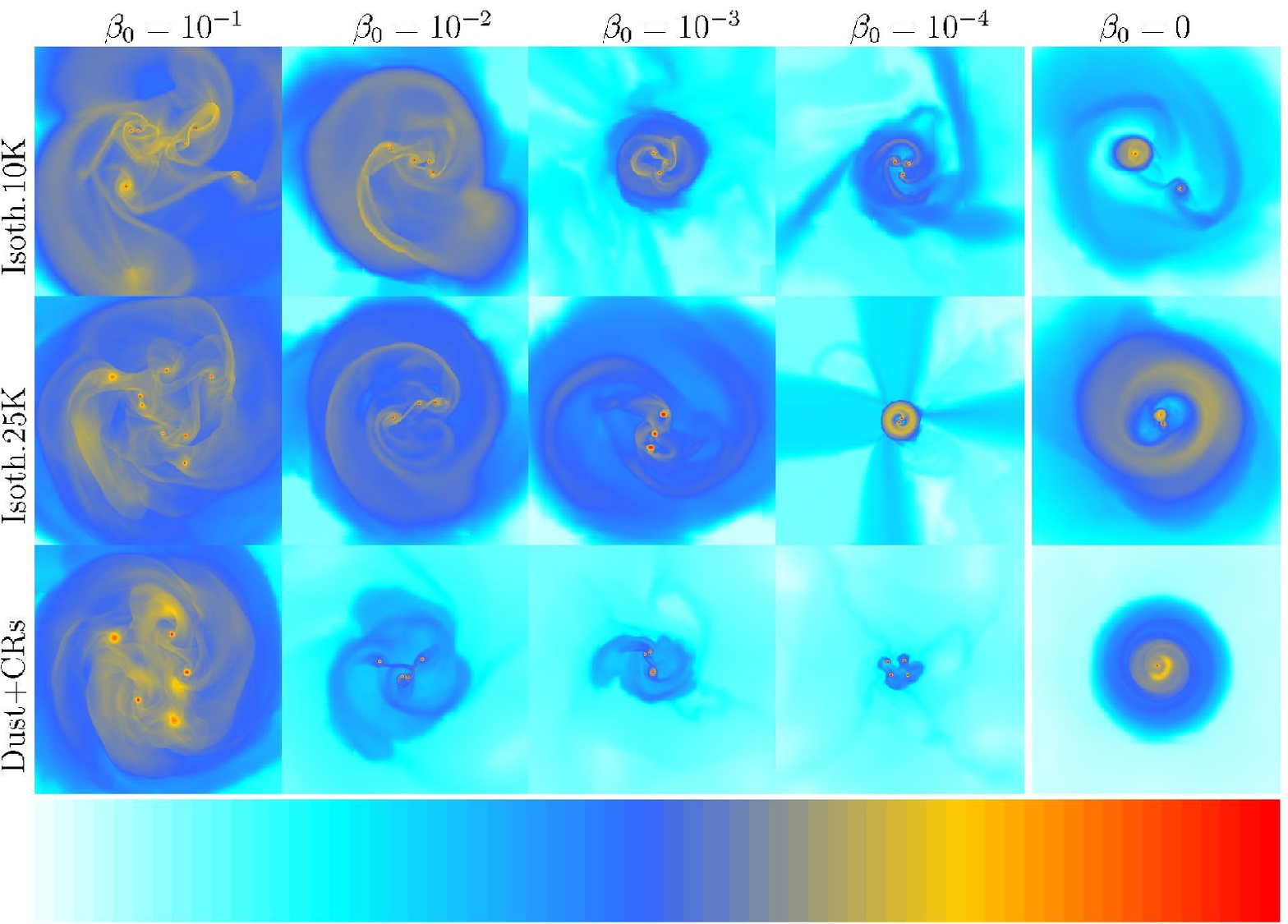}
\caption{Density plots of the environmental simulations at $t=12t_{\rm dyn}$. Molecular clouds in environments having dust and cosmic rays are simulated with solar metallicity. Orange to red depicts densities typically higher than $\rm 10^{4} ~cm^{-3}$ and up to $\rm 10^{9} ~cm^{-3}$. Light blue portrays the densities smaller than $\rm 1 ~cm^{-3}$. All images have boxsizes of 4x4 parsec.}
\label{fig:additional}
\end{figure*}

\begin{figure}[htb]
\centering
\includegraphics[scale=0.23]{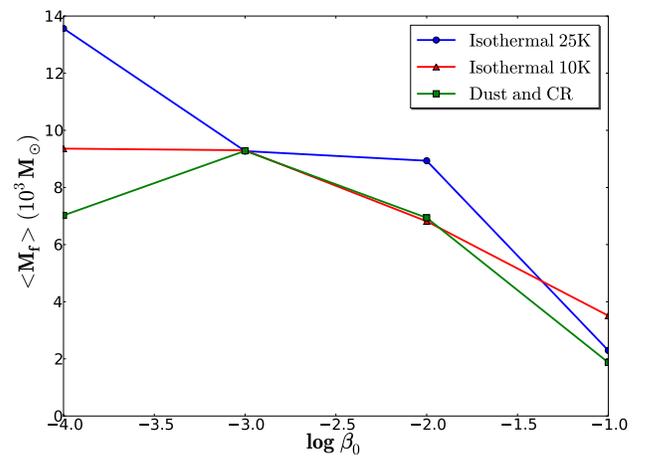}
\caption{Average fragment mass versus initial rotational energy for simulations within three distinct physical environments at $t = 12t_{\rm dyn}$.}
\label{fig:efficaddit}
\end{figure}

We present the density plots of the final stage of every run in figure \ref{fig:additional}. The results of these follow-up simulations can be found in table \ref{tab:fragments2} and the representative average fragment mass is plotted in figure \ref{fig:efficaddit}. \\

\begin{table*}[htb]
\begin{center}
\caption{Analysis of the fragments for the environmental simulations.}
\begin{tabular}{ccccc}
\hline \\
Simulation & Rot. energy & No. of fragments & Average mass & Individual fragment masses \\
$ $ & $\beta_{0}$ & $N_{f}$ & \textless$ M_{f} $\textgreater/$M_{\odot}$ & \textless$ M_{f} $\textgreater/$M_{\odot}$ \\
\\ \hline
\hline
Dust and CR	& $10^{-1}$	& 7	& 1873.9	& 454.4, 605.4, 1484.2, 1949.9, 2594.1, 2882.5, 3147.1\\
		& $10^{-2}$	& 4	& 6937.2	& 5732.3, 7075.8, 7417.4, 7523.5\\
		& $10^{-3}$	& 3	& 9282.3	& 6192.2, 6280.7, 15374.0\\
		& $10^{-4}$	& 4	& 7022.4	& 5707.4, 5727.1, 8286.7, 8368.3\\
		& $0$		& 1	& 11923.5	& No fragmentation\\
\hline
Isothermal 25K	& $10^{-1}$	& 8	& 2294.7	& 1407.6, 1696.0, 1981.3, 2096.3, 2301.5, 2317.8, 3067.5, 4262.3\\
		& $10^{-2}$	& 3	& 8935.7	& 6268.2, 8571.3, 11967.4\\
		& $10^{-3}$	& 3	& 9273.5	& 7537.8, 8878.8, 11404.0\\
		& $10^{-4}$	& 2	& 13565.8	& 11972.8, 15158.8\\
		& $0$		& 2	& 14754.1	& 9441.6, 20066.6\\
\hline
Isothermal 10K	& $10^{-1}$	& 7	& 3511.7	& 1801.5, 1886.3, 2041.8, 3881.1, 4372.6, 5126.8, 5471.8\\
		& $10^{-2}$	& 4	& 6815.6	& 4149.8, 6245.3, 6360.3, 10506.8\\
		& $10^{-3}$	& 3	& 9302.2	& 8452.4, 8939.4, 10514.8\\
		& $10^{-4}$	& 3	& 9360.1	& 5633.7, 11220.1, 11226.5\\
		& $0$		& 3	& 16381.3	& 9631.5, 23131.2\\
\hline
\end{tabular}
    \label{tab:fragments2}
\end{center}
\end{table*}

\textit{Dust and CRs in starbursts:} When we add an extra heating source due to dust, i.e., gas-grain collisional heating, and one due to elevated cosmic ray ionization, we see noticeable changes in the end results. Initially, the cooling becomes more efficient, because the gas has a higher temperature than the dust so that the dust acts as a coolant for the gas. When the gas temperature drops and falls below the dust temperature, the gas is heated by the dust (and cosmic rays). In the end, the fragment temperatures are close to the dust temperature of 39 K. The phase diagram for a non-rotating case in figure \ref{fig:phasediag-suppl} (upper left) shows this steep decrease in temperature, slightly overshooting below the dust temperature before it settles around the dust temperature. The strong collisional coupling between the dust and the gas imposes this tight $T-n$ relation.

Heating by cosmic rays and dust is enough to prevent further fragmentation to smaller scales and causes the largest fragments to accrete nearby material. Typically, fewer fragments are formed at this point, for the models with $\beta_0 \geq 10^{-2}$. There are more fragments for the case $\beta_0 = 10^{-4}$ where overall gas heating due to adiabatic compression and CRs renders $T_g > T_d$. This scenario is comparable to the situation for population II.5 stars \citep{2006MNRAS.369.1437S}. So, dust can enhance fragmentation when the conditions are right.

A remarkable result that we see is that when there is zero initial rotation, the warm dusty cloud stops fragmenting in contrast to the expected trend. The would-be fragments are blurred out and settle in a disk around the sole compact object instead. We find that the cloud is very sensitive to initial rotation in this regime. A rapid turnover from a fragmenting cloud to a non-fragmenting case is found to be around $\beta_0 \sim 10^{-5}$. Interestingly, this is in very good agreement with the lower limit of fragmentation ($\beta_0 = 10^{-4}$) that \cite{2009arXiv0907.3257M} find. Starburst systems may have super-solar metallicities \citep{2009A&A...493..525I}. We have verified that increasing the metallicity of the cloud to twice solar has little impact on our results. \\

\textit{Isothermal:} When we keep the temperature constant, we effectively assume that the pressure scales linearly with the density. Two near-isothermal runs are compared to the thermal balance runs at 10 K and 25 K. In environments that are less active, molecular clouds are known to have these low temperatures. Although we initialize with these temperatures, accretion shocks, that can increase the temperature up to a few hundred degrees in their wake, still affect the temperature of the cloud and the cores. Typically, for gas with density higher than $n=1 \rm ~cm^{-3}$, the temperature remains constant around the initialized value, see the two diagrams on the right in figure \ref{fig:phasediag-suppl}, while the temperature for the diffuse gas can go up one order of magnitude. For the 25 K condition, this makes the Jeans mass higher compared to simulations with incorporation of the thermal balance and where the cooling is dominant. Compared to the highest metallicity case $Z=Z_{\odot}$ of the metallicity study, for $\beta_{0}\gtrsim10^{-3}$, the fragments are less well developed and have densities similar to the lower metallicity results. Even for the 10 K isothermal case, though it has the same final temperature as the cooling included runs, there are fewer fragments.

At lower rotational energies the difference is less apparent, since compressional heating for non-isothermal runs can overcome the cooling. In fact, for non-rotating clouds, fragmentation for the 10 K isothermal case tends to be higher. This change in fragmentation is caused by the fact that there is a temperature gradient with increasing density. Although the thermal Jeans mass of the metallicity runs is initially higher than in all the near-isothermal models, fragmentation is stimulated due to softening of $\gamma$. We note that for the lowest rotational energy ($\beta_{0}=10^{-4}$) and the non-rotating case in the solar metallicity model, $\gamma$ is stiff ($\gamma=1.4$), and the temperature stays relatively high ($\gtrsim$100 K) compared to the near-isothermal case, because compressional heating suppresses fragmentation. The evident difference between the two near-isothermal cases (10 and 25 K) can be explained by the difference in Jeans mass of a factor $\sim$4.

\section{Conclusions and Discussion}
We have performed 38 hydrodynamical simulations where we have rigorously tested the dependence of molecular cloud fragmentation on the ambient conditions as they pertain to starburst and dwarf galaxy regions. We have compared these with the well known conditions of the Milky Way. Our results on the metallicity simulations show that the amount of fragmentation and the compressibility of the gas scales with increasing metallicity. In this, the cooling by gas phase and grain surface chemical species plays a subtle but fundamental role. We find that turbulence and rotation also play an important role. Turbulence of 0.3$E_{\rm grav}$, following the Larson laws, is a crucial ingredient in the development of fragments. Although rotation is observed for sub-pc scale molecular clouds, it is unlikely that clouds of these sizes (parsec) have rotational support. All simulations have therefore been performed for $\beta_0$ values of much less than unity. These results agree with expectations and are comparable to the work of \cite{Machida1} and \cite{2009arXiv0907.3257M}, albeit at lower density. The metallicity has a greater impact on cloud fragmentation than $\beta_0$. For the highest metallicity model, fragmentation (into a binary) is still possible for even the lowest (or zero) rotational energy.

Regarding the initial temperature of the simulations, we note that starting with a high temperature of 100 K, as for starbursts, does not seem to be too important. When we do not consider the extra heating sources, radiative cooling can quickly overcome the high initial temperature and cool the cloud down to $\sim$10 K. This is true for most cases except when the initial rotational energy is very low. In this situation, the adiabatic heating and the heating due to shocks, which are more prominent now, can overcome the cooling. Fragments have higher temperatures $>$100 K under these circumstances, which suppresses fragmentation. The surrounding gas, though, stays cold ($\sim$10-20 K).

When we use low metallicity, similar to dwarf galaxies or the early universe, the average fragment mass increases in our simulations. This leads to massive cloud cores and eventually can induce massive star formation \citep{2009arXiv0901.0915D, 2008ApJ...672..410L}, relevant to the origin of the IMF. \\

We summarize our main conclusions as follows:
\begin{itemize}
\item Increased metallicity promotes fragmentation. There is a clear and strong dependence of fragmentation on metallicity, where the average fragment mass decreases with increasing metallicity.

\item Heating by cosmic rays and dust, for high metallicity initial conditions, are strong enough to slow down cooling and reduce fragmentation. When the gas and dust are thermally coupled, we find that the gas temperature stabilizes close to the dust temperature, T$_g$ = T$_d$, at densities $>$10$^5 \rm ~cm^{-3}$.

\item Fragmentation is suppressed for isothermal clouds. In almost all of the near-isothermal simulations that we did, we see a smaller number of fragments compared to their non-isothermal counterparts, except when adiabatic heating is very strong. Although the temperatures are exactly the same in several cases, even lower compared to the non-isothermal runs for the smallest $\beta_{0}$ ratios, and the Jeans masses are comparable, the evolution and fragmentation of the cloud differs. We attribute this to the stiffer effective equation of state, i.e. $\gamma=1$, that evidently has a great impact \citep{2005IAUS..227..337K}.
\end{itemize}

For the main part in this research, we have performed 2D simulations. One can argue that 2D simulations are not representative for a 3D cloud. Certain physics, like MHD, is indeed not possible to do in 2D, but we have not used any kind of physics or initial condition that requires a third dimension. We find that a higher merger rate occurs for 2D simulations due to one less degree of freedom and that our fragments are of somewhat higher mass. Still, we do not see many mergers in general. The 3D simulations that we have run, show that a disk forms in a time that is comparable to or is less than the fragmentation time scale if even a modest rotation exists, effectively making the problem a two-dimensional one. The results are slightly different, perhaps due to the lower resolution, or possibly owing to the shorter simulation time, but show comparable structures. A disk does not form when there is complete lack of initial rotation. This, however, does not change our results, since the non-rotating 3D simulation also fragments into two compact objects with comparable masses and enjoys a similar temperature evolution (figure \ref{fig:phasediag-suppl}) as its 2D counterpart.

Changes in the density profile do not affect the evolution of the cloud as far as the impact of metallicity is concerned. Testing a power-law density profile (eq. \ref{eq:profile}), as observed in active galactic regions \citep{2003ApJ...594..812G, 2008arXiv0805.0185M}, against a flat profile and keeping the mean density constant, we find that the results are similar. The results (not shown) are the same for the number of fragments, but also alike in their temperature evolution. Only the fragment masses are marginally smaller when using a flat profile, due to competitive accretion \citep{2005ASSL..327..425B}.

\begin{equation}
\rho = \rho_0 \left( \frac{1} {1 + R/R_c} \right)^{\alpha} $~~cm$^{-3}.
\label{eq:profile}
\end{equation}

\noindent
$\rho_0$ is the central density of $n=10^4 \rm ~cm^{-3}$, R$_{c}$ is the characteristic length of 1 pc. This is the radius within which the central density remains approximately constant. $R$ is the specific cloud radius and $\alpha=1.4$ is the power of the profile. \\

We do not reach star densities, but the fragmentation and formation of dense cores is determined at the much earlier phases that we probe. We have not made use of sink particles, as this is usually done to make a jump in resolution and form protostars from dense cores, but our fragments should be able to form many stars. Opacity effects beyond 10$^{22.5} \rm ~cm^{-2}$ are not considered in our simulations simply because we do not reach very high densities. The maximum densities that we reach are n$\sim$10$^8$-10$^9$ cm$^{-3}$. Molecular opacity at high densities can actually have a quite different effect on fragmentation compared to densities below 10$^{3} \rm ~cm^{-3}$ \citep{2005A&A...440..559P, 2006A&A...453..615P}. A cloud can become optically thick sooner and trap the lines when there are more metals to form molecules with. This in turn will heat the system and prevent further fragmentation. Incorporating radiative transfer will be the next step in the continuation of this research.

\begin{acknowledgements}
We would like to thank Floris van der Tak for useful discussions. Special thanks goes to Latif for all the long and fruitful conversations. We are very grateful to the anonymous referee for an insightful and constructive report that greatly helped this work. The FLASH code was in part developed by the DOE-supported Alliance Center for Astrophysical Thermonuclear Flashes (ACS) at the University of Chicago. Our simulations were carried out on the Donald Smits Center for Information Technology using the HPC Cluster, University of Groningen. \\
\end{acknowledgements}

\bibliography{biblio.future.bib}

\begin{thebibliography}{85}
\expandafter\ifx\csname natexlab\endcsname\relax\def\natexlab#1{#1}\fi

\bibitem[{{Abel} {et~al.}(2000){Abel}, {Bryan}, \&
  {Norman}}]{2000ApJ...540...39A}
{Abel}, T., {Bryan}, G.~L., \& {Norman}, M.~L. 2000, \apj, 540, 39

\bibitem[{{Abel} {et~al.}(2002){Abel}, {Bryan}, \&
  {Norman}}]{2002Sci...295...93A}
{Abel}, T., {Bryan}, G.~L., \& {Norman}, M.~L. 2002, Science, 295, 93

\bibitem[{{Alves} {et~al.}(2001){Alves}, {Lada}, \&
  {Lada}}]{2001Natur.409..159A}
{Alves}, J.~F., {Lada}, C.~J., \& {Lada}, E.~A. 2001, \nat, 409, 159

\bibitem[{{Ballero} {et~al.}(2007){Ballero}, {Kroupa}, \&
  {Matteucci}}]{2007A&A...467..117B}
{Ballero}, S.~K., {Kroupa}, P., \& {Matteucci}, F. 2007, \aap, 467, 117

\bibitem[{{Ballesteros-Paredes} {et~al.}(1999){Ballesteros-Paredes},
  {Hartmann}, \& {V{\'a}zquez-Semadeni}}]{1999ApJ...527..285B}
{Ballesteros-Paredes}, J., {Hartmann}, L., \& {V{\'a}zquez-Semadeni}, E. 1999,
  \apj, 527, 285

\bibitem[{{Banerjee} {et~al.}(2007){Banerjee}, {Klessen}, \&
  {Fendt}}]{2007ApJ...668.1028B}
{Banerjee}, R., {Klessen}, R.~S., \& {Fendt}, C. 2007, \apj, 668, 1028

\bibitem[{{Banerjee} {et~al.}(2008){Banerjee}, {Vazquez-Semadeni},
  {Hennebelle}, \& {Klessen}}]{2008arXiv0808.0986B}
{Banerjee}, R., {Vazquez-Semadeni}, E., {Hennebelle}, P., \& {Klessen}, R.
  2008, ArXiv e-prints, 808

\bibitem[{{Bodenheimer}(1995)}]{1995ARA&A..33..199B}
{Bodenheimer}, P. 1995, \araa, 33, 199

\bibitem[{{Bonnell}(2005)}]{2005ASSL..327..425B}
{Bonnell}, I.~A. 2005, in Astrophysics and Space Science Library, Vol. 327, The
  Initial Mass Function 50 Years Later, ed. E.~{Corbelli}, F.~{Palla}, \&
  H.~{Zinnecker}, 425--+

\bibitem[{{Bonnell} {et~al.}(2006){Bonnell}, {Dobbs}, {Robitaille}, \&
  {Pringle}}]{2006MNRAS.365...37B}
{Bonnell}, I.~A., {Dobbs}, C.~L., {Robitaille}, T.~P., \& {Pringle}, J.~E.
  2006, \mnras, 365, 37

\bibitem[{{Bromm} {et~al.}(2002){Bromm}, {Coppi}, \&
  {Larson}}]{2002ApJ...564...23B}
{Bromm}, V., {Coppi}, P.~S., \& {Larson}, R.~B. 2002, \apj, 564, 23

\bibitem[{{Caselli} {et~al.}(2002){Caselli}, {Benson}, {Myers}, \&
  {Tafalla}}]{2002ApJ...572..238C}
{Caselli}, P., {Benson}, P.~J., {Myers}, P.~C., \& {Tafalla}, M. 2002, \apj,
  572, 238

\bibitem[{{Caselli} {et~al.}(2008){Caselli}, {Vastel}, {Ceccarelli}, {van der
  Tak}, {Crapsi}, \& {Bacmann}}]{2008A&A...492..703C}
{Caselli}, P., {Vastel}, C., {Ceccarelli}, C., {et~al.} 2008, \aap, 492, 703

\bibitem[{{Cazaux} {et~al.}(2005){Cazaux}, {Caselli}, {Tielens}, {LeBourlot},
  \& {Walmsley}}]{2005JPhCS...6..155C}
{Cazaux}, S., {Caselli}, P., {Tielens}, A.~G.~G.~M., {LeBourlot}, J., \&
  {Walmsley}, M. 2005, Journal of Physics Conference Series, 6, 155

\bibitem[{{Cazaux} \& {Spaans}(2004)}]{2004ApJ...611...40C}
{Cazaux}, S. \& {Spaans}, M. 2004, \apj, 611, 40

\bibitem[{{Cazaux} \& {Spaans}(2009)}]{2009A&A...496..365C}
{Cazaux}, S. \& {Spaans}, M. 2009, \aap, 496, 365

\bibitem[{Childs {et~al.}(2005)Childs, Brugger, Bonnell, Meredith, Miller,
  Whitlock, \& Max}]{Childs:2005:ACS}
Childs, H., Brugger, E.~S., Bonnell, K.~S., {et~al.} 2005, in Proceedings of
  IEEE Visualization 2005, 190--198

\bibitem[{{Clark} {et~al.}(2005){Clark}, {Bonnell}, {Zinnecker}, \&
  {Bate}}]{2005MNRAS.359..809C}
{Clark}, P.~C., {Bonnell}, I.~A., {Zinnecker}, H., \& {Bate}, M.~R. 2005,
  \mnras, 359, 809

\bibitem[{{Clark} {et~al.}(2008){Clark}, {Glover}, \&
  {Klessen}}]{2008ApJ...672..757C}
{Clark}, P.~C., {Glover}, S.~C.~O., \& {Klessen}, R.~S. 2008, \apj, 672, 757

\bibitem[{{Colella} \& {Woodward}(1984)}]{1984JCoPh..54..174C}
{Colella}, P. \& {Woodward}, P.~R. 1984, Journal of Computational Physics, 54,
  174

\bibitem[{{Commercon} {et~al.}(2007){Commercon}, {Hennebelle}, {Audit},
  {Chabrier}, \& {Teyssier}}]{2007arXiv0709.2450C}
{Commercon}, B., {Hennebelle}, P., {Audit}, E., {Chabrier}, G., \& {Teyssier},
  R. 2007, ArXiv e-prints, 709

\bibitem[{{Cunha} {et~al.}(2008){Cunha}, {Smith}, {Sellgren}, {Blum},
  {Ram{\'{\i}}rez}, \& {Terndrup}}]{2008IAUS..245..339C}
{Cunha}, K., {Smith}, V.~V., {Sellgren}, K., {et~al.} 2008, in IAU Symposium,
  Vol. 245, IAU Symposium, 339--342

\bibitem[{{Dabringhausen} {et~al.}(2009){Dabringhausen}, {Kroupa}, \&
  {Baumgardt}}]{2009arXiv0901.0915D}
{Dabringhausen}, J., {Kroupa}, P., \& {Baumgardt}, H. 2009, ArXiv
  e-prints:0901.0915

\bibitem[{{Dib} {et~al.}(2007){Dib}, {Kim}, \&
  {Shadmehri}}]{2007MNRAS.381L..40D}
{Dib}, S., {Kim}, J., \& {Shadmehri}, M. 2007, \mnras, 381, L40

\bibitem[{{Dobbs}(2008)}]{2008MNRAS.391..844D}
{Dobbs}, C.~L. 2008, \mnras, 391, 844

\bibitem[{{Dobbs} \& {Bonnell}(2008)}]{2008MNRAS.385.1893D}
{Dobbs}, C.~L. \& {Bonnell}, I.~A. 2008, \mnras, 385, 1893

\bibitem[{{Dulieu} {et~al.}(2009){Dulieu}, {Amiaud}, {Congiu}, {Fillion},
  {Matar}, {Momeni}, {Pirronello}, \& {Lemaire}}]{2009arXiv0903.3120D}
{Dulieu}, F., {Amiaud}, L., {Congiu}, E., {et~al.} 2009, ArXiv
  e-prints:0903.3120

\bibitem[{{Elmegreen}(2000)}]{2000ApJ...539..342E}
{Elmegreen}, B.~G. 2000, \apj, 539, 342

\bibitem[{{Elmegreen} {et~al.}(2008){Elmegreen}, {Klessen}, \&
  {Wilson}}]{2008ApJ...681..365E}
{Elmegreen}, B.~G., {Klessen}, R.~S., \& {Wilson}, C.~D. 2008, \apj, 681, 365

\bibitem[{{Falgarone} {et~al.}(2001){Falgarone}, {Pety}, \&
  {Phillips}}]{2001ApJ...555..178F}
{Falgarone}, E., {Pety}, J., \& {Phillips}, T.~G. 2001, \apj, 555, 178

\bibitem[{{Figer}(2005{\natexlab{a}})}]{2005Natur.434..192F}
{Figer}, D.~F. 2005{\natexlab{a}}, \nat, 434, 192

\bibitem[{{Figer}(2005{\natexlab{b}})}]{2005ASSL..327...89F}
{Figer}, D.~F. 2005{\natexlab{b}}, in Astrophysics and Space Science Library,
  Vol. 327, The Initial Mass Function 50 Years Later, ed. E.~{Corbelli},
  F.~{Palla}, \& H.~{Zinnecker}, 89--+

\bibitem[{{Figer} {et~al.}(1999){Figer}, {Morris}, {Kim}, \&
  {Serabyn}}]{1999ASPC..186..329F}
{Figer}, D.~F., {Morris}, M., {Kim}, S.~S., \& {Serabyn}, E. 1999, in
  Astronomical Society of the Pacific Conference Series, Vol. 186, The Central
  Parsecs of the Galaxy, ed. H.~{Falcke}, A.~{Cotera}, W.~J. {Duschl},
  F.~{Melia}, \& M.~J. {Rieke}, 329--+

\bibitem[{{Frieswijk}(2008)}]{Frieswijk2008}
{Frieswijk}, W.~W.~F. 2008, PhD thesis, Kapteyn Astronomical Institute,
  University of Groningen

\bibitem[{{Fryxell} {et~al.}(2000){Fryxell}, {Olson}, {Ricker}, {Timmes},
  {Zingale}, {Lamb}, {MacNeice}, {Rosner}, {Truran}, \&
  {Tufo}}]{2000ApJS..131..273F}
{Fryxell}, B., {Olson}, K., {Ricker}, P., {et~al.} 2000, \apjs, 131, 273

\bibitem[{{Genzel} {et~al.}(2003){Genzel}, {Sch{\"o}del}, {Ott}, {Eisenhauer},
  {Hofmann}, {Lehnert}, {Eckart}, {Alexander}, {Sternberg}, {Lenzen},
  {Cl{\'e}net}, {Lacombe}, {Rouan}, {Renzini}, \&
  {Tacconi-Garman}}]{2003ApJ...594..812G}
{Genzel}, R., {Sch{\"o}del}, R., {Ott}, T., {et~al.} 2003, \apj, 594, 812

\bibitem[{{Goodwin} {et~al.}(2008){Goodwin}, {Nutter}, {Kroupa},
  {Ward-Thompson}, \& {Whitworth}}]{2008A&A...477..823G}
{Goodwin}, S.~P., {Nutter}, D., {Kroupa}, P., {Ward-Thompson}, D., \&
  {Whitworth}, A.~P. 2008, \aap, 477, 823

\bibitem[{{Greif} {et~al.}(2008){Greif}, {Schleicher}, {Johnson}, {Jappsen},
  {Klessen}, {Clark}, {Glover}, {Stacy}, \& {Bromm}}]{2008IAUS..255...33G}
{Greif}, T.~H., {Schleicher}, D.~R.~G., {Johnson}, J.~L., {et~al.} 2008, in IAU
  Symposium, Vol. 255, IAU Symposium, 33--48

\bibitem[{{Habing}(1968)}]{1968BAN....19..421H}
{Habing}, H.~J. 1968, \bain, 19, 421

\bibitem[{{Hartmann} {et~al.}(2001){Hartmann}, {Ballesteros-Paredes}, \&
  {Bergin}}]{2001ApJ...562..852H}
{Hartmann}, L., {Ballesteros-Paredes}, J., \& {Bergin}, E.~A. 2001, \apj, 562,
  852

\bibitem[{{Heitsch} \& {Hartmann}(2008)}]{2008ApJ...689..290H}
{Heitsch}, F. \& {Hartmann}, L. 2008, \apj, 689, 290

\bibitem[{{Hollenbach} \& {McKee}(1979)}]{1979ApJS...41..555H}
{Hollenbach}, D. \& {McKee}, C.~F. 1979, \apjs, 41, 555

\bibitem[{{Hollenbach} \& {McKee}(1989)}]{1989ApJ...342..306H}
{Hollenbach}, D. \& {McKee}, C.~F. 1989, \apj, 342, 306

\bibitem[{{Hunter}(2008)}]{2008IAUS..255..226H}
{Hunter}, D.~A. 2008, in IAU Symposium, Vol. 255, IAU Symposium, 226--237

\bibitem[{{Israel}(2009)}]{2009A&A...493..525I}
{Israel}, F.~P. 2009, \aap, 493, 525

\bibitem[{{Jappsen} {et~al.}(2008){Jappsen}, {Mac Low}, {Glover}, {Klessen}, \&
  {Kitsionas}}]{2008arXiv0810.1867J}
{Jappsen}, A.~., {Mac Low}, M.~., {Glover}, S.~C.~O., {Klessen}, R.~S., \&
  {Kitsionas}, S. 2008, ArXiv e-prints:0810.1867

\bibitem[{{Kim} {et~al.}(2006){Kim}, {Figer}, {Kudritzki}, \&
  {Najarro}}]{2006ApJ...653L.113K}
{Kim}, S.~S., {Figer}, D.~F., {Kudritzki}, R.~P., \& {Najarro}, F. 2006, \apjl,
  653, L113

\bibitem[{{Klessen} {et~al.}(2005{\natexlab{a}}){Klessen},
  {Ballesteros-Paredes}, {V{\'a}zquez-Semadeni}, \&
  {Dur{\'a}n-Rojas}}]{2005ApJ...620..786K}
{Klessen}, R.~S., {Ballesteros-Paredes}, J., {V{\'a}zquez-Semadeni}, E., \&
  {Dur{\'a}n-Rojas}, C. 2005{\natexlab{a}}, \apj, 620, 786

\bibitem[{{Klessen} {et~al.}(2005{\natexlab{b}}){Klessen}, {Spaans}, \&
  {Jappsen}}]{2005IAUS..227..337K}
{Klessen}, R.~S., {Spaans}, M., \& {Jappsen}, A.-K. 2005{\natexlab{b}}, in IAU
  Symposium, Vol. 227, Massive Star Birth: A Crossroads of Astrophysics, ed.
  R.~{Cesaroni}, M.~{Felli}, E.~{Churchwell}, \& M.~{Walmsley}, 337--345

\bibitem[{{Klessen} {et~al.}(2007){Klessen}, {Spaans}, \&
  {Jappsen}}]{2007MNRAS.374L..29K}
{Klessen}, R.~S., {Spaans}, M., \& {Jappsen}, A.-K. 2007, \mnras, 374, L29

\bibitem[{{Lada} {et~al.}(2008){Lada}, {Muench}, {Rathborne}, {Alves}, \&
  {Lombardi}}]{2008ApJ...672..410L}
{Lada}, C.~J., {Muench}, A.~A., {Rathborne}, J., {Alves}, J.~F., \& {Lombardi},
  M. 2008, \apj, 672, 410

\bibitem[{{Larson}(1981)}]{1981MNRAS.194..809L}
{Larson}, R.~B. 1981, \mnras, 194, 809

\bibitem[{{Machida}(2008)}]{Machida1}
{Machida}, M.~N. 2008, ArXiv e-prints, 802

\bibitem[{{Machida} {et~al.}(2009){Machida}, {Omukai}, {Matsumoto}, \&
  {Inutsuka}}]{2009arXiv0907.3257M}
{Machida}, M.~N., {Omukai}, K., {Matsumoto}, T., \& {Inutsuka}, S.-i. 2009,
  ArXiv e-prints

\bibitem[{{MacNeice} {et~al.}(2000){MacNeice}, {Olson}, {Mobarry}, {de
  Fainchtein}, \& {Packer}}]{2000CoPhC.126..330M}
{MacNeice}, P., {Olson}, K.~M., {Mobarry}, C., {de Fainchtein}, R., \&
  {Packer}, C. 2000, Computer Physics Communications, 126, 330

\bibitem[{{Mapelli} {et~al.}(2008){Mapelli}, {Hayfield}, {Mayer}, \&
  {Wadsley}}]{2008arXiv0805.0185M}
{Mapelli}, M., {Hayfield}, T., {Mayer}, L., \& {Wadsley}, J. 2008, ArXiv
  e-prints, 805

\bibitem[{{Mathis} {et~al.}(1977){Mathis}, {Rumpl}, \&
  {Nordsieck}}]{1977ApJ...217..425M}
{Mathis}, J.~S., {Rumpl}, W., \& {Nordsieck}, K.~H. 1977, \apj, 217, 425

\bibitem[{{Matsumoto} \& {Hanawa}(2003)}]{2003ApJ...595..913M}
{Matsumoto}, T. \& {Hanawa}, T. 2003, \apj, 595, 913

\bibitem[{{Meijerink} \& {Spaans}(2005)}]{2005A&A...436..397M}
{Meijerink}, R. \& {Spaans}, M. 2005, \aap, 436, 397

\bibitem[{{Motte} \& {Andr{\'e}}(2001)}]{2001ASPC..243..301M}
{Motte}, F. \& {Andr{\'e}}, P. 2001, in Astronomical Society of the Pacific
  Conference Series, Vol. 243, From Darkness to Light: Origin and Evolution of
  Young Stellar Clusters, ed. T.~{Montmerle} \& P.~{Andr{\'e}}, 301--+

\bibitem[{{Motte} {et~al.}(1998){Motte}, {Andre}, \&
  {Neri}}]{1998A&A...336..150M}
{Motte}, F., {Andre}, P., \& {Neri}, R. 1998, \aap, 336, 150

\bibitem[{{Olson} {et~al.}(1999){Olson}, {MacNeice}, {Fryxell}, {Ricker},
  {Timmes}, \& {Zingale}}]{1999AAS...195.4203O}
{Olson}, K.~M., {MacNeice}, P., {Fryxell}, B., {et~al.} 1999, in Bulletin of
  the American Astronomical Society, Vol.~31, Bulletin of the American
  Astronomical Society, 1430--+

\bibitem[{{Omukai} \& {Palla}(2001)}]{2001ApJ...561L..55O}
{Omukai}, K. \& {Palla}, F. 2001, \apjl, 561, L55

\bibitem[{{Omukai} \& {Palla}(2003)}]{2003ApJ...589..677O}
{Omukai}, K. \& {Palla}, F. 2003, \apj, 589, 677

\bibitem[{{Omukai} {et~al.}(2005){Omukai}, {Tsuribe}, {Schneider}, \&
  {Ferrara}}]{2005ApJ...626..627O}
{Omukai}, K., {Tsuribe}, T., {Schneider}, R., \& {Ferrara}, A. 2005, \apj, 626,
  627

\bibitem[{{Ott} {et~al.}(2005){Ott}, {Weiss}, {Henkel}, \&
  {Walter}}]{2005ApJ...629..767O}
{Ott}, J., {Weiss}, A., {Henkel}, C., \& {Walter}, F. 2005, \apj, 629, 767

\bibitem[{{Poelman} \& {Spaans}(2005)}]{2005A&A...440..559P}
{Poelman}, D.~R. \& {Spaans}, M. 2005, \aap, 440, 559

\bibitem[{{Poelman} \& {Spaans}(2006)}]{2006A&A...453..615P}
{Poelman}, D.~R. \& {Spaans}, M. 2006, \aap, 453, 615

\bibitem[{{Portegies Zwart} {et~al.}(2007){Portegies Zwart}, {Gaburov}, {Chen},
  \& {G{\"u}rkan}}]{2007MNRAS.378L..29P}
{Portegies Zwart}, S., {Gaburov}, E., {Chen}, H.-C., \& {G{\"u}rkan}, M.~A.
  2007, \mnras, 378, L29

\bibitem[{{Sabbi} {et~al.}(2007){Sabbi}, {Sirianni}, {Nota}, {Gallagher},
  {Tosi}, {Smith}, {Angeretti}, \& {Meixner}}]{2007AAS...210.1601S}
{Sabbi}, E., {Sirianni}, M., {Nota}, A., {et~al.} 2007, in American
  Astronomical Society Meeting Abstracts, Vol. 210, American Astronomical
  Society Meeting Abstracts, 16.01--+

\bibitem[{{Salvadori} \& {Ferrara}(2009)}]{2009MNRAS.tmpL.207S}
{Salvadori}, S. \& {Ferrara}, A. 2009, \mnras, L207+

\bibitem[{{Schneider} {et~al.}(2006){Schneider}, {Omukai}, {Inoue}, \&
  {Ferrara}}]{2006MNRAS.369.1437S}
{Schneider}, R., {Omukai}, K., {Inoue}, A.~K., \& {Ferrara}, A. 2006, \mnras,
  369, 1437

\bibitem[{{Simpson} {et~al.}(2008){Simpson}, {Nutter}, \&
  {Ward-Thompson}}]{2008MNRAS.391..205S}
{Simpson}, R.~J., {Nutter}, D., \& {Ward-Thompson}, D. 2008, \mnras, 391, 205

\bibitem[{{Skillman} {et~al.}(2003){Skillman}, {Tolstoy}, {Cole}, {Dolphin},
  {Saha}, {Gallagher}, {Dohm-Palmer}, \& {Mateo}}]{2003ApJ...596..253S}
{Skillman}, E.~D., {Tolstoy}, E., {Cole}, A.~A., {et~al.} 2003, \apj, 596, 253

\bibitem[{{Smith} {et~al.}(2009){Smith}, {Clark}, \&
  {Bonnell}}]{2009arXiv0903.3240S}
{Smith}, R.~J., {Clark}, P.~C., \& {Bonnell}, I.~A. 2009, ArXiv e-prints

\bibitem[{{Spaans} \& {Norman}(1997)}]{1997ApJ...483...87S}
{Spaans}, M. \& {Norman}, C.~A. 1997, \apj, 483, 87

\bibitem[{{Spaans} \& {Silk}(2000)}]{2000ApJ...538..115S}
{Spaans}, M. \& {Silk}, J. 2000, \apj, 538, 115

\bibitem[{{Spaans} \& {Silk}(2005)}]{2005ApJ...626..644S}
{Spaans}, M. \& {Silk}, J. 2005, \apj, 626, 644

\bibitem[{{Stolte} {et~al.}(2005){Stolte}, {Brandner}, {Grebel}, {Lenzen}, \&
  {Lagrange}}]{2005ApJ...628L.113S}
{Stolte}, A., {Brandner}, W., {Grebel}, E.~K., {Lenzen}, R., \& {Lagrange},
  A.-M. 2005, \apjl, 628, L113

\bibitem[{{Stolte} {et~al.}(2002){Stolte}, {Grebel}, {Brandner}, \&
  {Figer}}]{2002A&A...394..459S}
{Stolte}, A., {Grebel}, E.~K., {Brandner}, W., \& {Figer}, D.~F. 2002, \aap,
  394, 459

\bibitem[{{Tilley} \& {Pudritz}(2005)}]{2005JRASC..99R.132T}
{Tilley}, D. \& {Pudritz}, R. 2005, \jrasc, 99, 132

\bibitem[{{Tilley} \& {Pudritz}(2007)}]{2007MNRAS.382...73T}
{Tilley}, D.~A. \& {Pudritz}, R.~E. 2007, \mnras, 382, 73

\bibitem[{{Truelove} {et~al.}(1997){Truelove}, {Klein}, {McKee}, {Holliman},
  {Howell}, \& {Greenough}}]{1997ApJ...489L.179T}
{Truelove}, J.~K., {Klein}, R.~I., {McKee}, C.~F., {et~al.} 1997, \apjl, 489,
  L179+

\bibitem[{{Wiklind} \& {Henkel}(2001)}]{2001A&A...375..797W}
{Wiklind}, T. \& {Henkel}, C. 2001, \aap, 375, 797

\bibitem[{{Yoshida} {et~al.}(2006){Yoshida}, {Omukai}, {Hernquist}, \&
  {Abel}}]{2006ApJ...652....6Y}
{Yoshida}, N., {Omukai}, K., {Hernquist}, L., \& {Abel}, T. 2006, \apj, 652, 6

\end{thebibliography}


\end{document}